%% file: LoopsIndS_PRD.tex
\documentclass[12pt]{article}
\pdfoutput=1

\usepackage{cite}
\usepackage{booktabs}
\usepackage[english]{babel}
\usepackage{amsmath,amssymb,amsbsy,amstext, amsthm, simplewick}
\usepackage{hyperref}
\usepackage{graphicx}
\usepackage{amsfonts}
\usepackage{amssymb}
\usepackage[small]{caption}
\usepackage{upgreek}
\usepackage[svgnames,dvipsnames,x11names,table]{xcolor}
\usepackage{multirow}
\usepackage{geometry}
\usepackage[hang,flushmargin]{footmisc}
\usepackage{bm}
\usepackage{braket}
\usepackage{subcaption}
\usepackage{mathtools}
\usepackage{setspace}
\usepackage{cleveref}
\usepackage{comment}
\usepackage{scalerel}
\usepackage[normalem]{ulem}
\usepackage{slashed}
\usepackage{enumitem}
\usepackage{ifthen}
\usepackage{appendix}
\usepackage{floatrow}
\usepackage{tikz}
\usetikzlibrary{decorations.markings}
\usetikzlibrary{shapes.misc}
\usetikzlibrary{arrows}

\makeatletter
\g@addto@macro\bfseries{\boldmath}
\makeatother

\hypersetup{
    colorlinks=true,
    linkcolor={red!50!black},
    citecolor={blue!50!black},
    urlcolor={blue!80!black}
}

\usepackage{colortbl}

\makeatletter
\newlength{\apb@width}
\newcommand{\autoparbox}[2][c]{\settowidth{\apb@width}{#2}\parbox[#1]{\apb@width}{#2}}

\makeatother

\definecolor{lightgray}{gray}{0.9}

\usepackage[framemethod=default]{mdframed}
\newmdenv[skipabove=7pt,
skipbelow=7pt,
rightline=false,
leftline=false,
topline=false,
bottomline=false,
backgroundcolor=gray!10,
linecolor=gray,
innerleftmargin=5pt,
innerrightmargin=5pt,
innertopmargin=5pt,
innerbottommargin=5pt,
leftmargin=0cm,
rightmargin=0cm,
linewidth=4pt]{eBox}

\usepackage[most]{tcolorbox}
\tcbset{colback=white, colframe=black,
        highlight math style= {enhanced, 
            colframe=red,colback=red!10!white,boxsep=0pt}
        }
\definecolor{light-gray}{gray}{0.95}

\crefname{table}{Table}{Tables}
\crefname{equation}{Eq.}{Eqs.}
\crefname{appendix}{App.}{Apps.}
\crefname{section}{Sec.}{Secs.}
\crefname{figure}{Fig.}{Figs.}

\numberwithin{equation}{section}

\def \d {{\mathrm{d}}}
\def \e {{\epsilon}}

\def \i {{i}}

\def \x {{\vec x}}
\def \y {{\vec y}}
\def \z {{\vec z}}

\def \k {{\vec k}}
\def \p {{\vec p}}

\def \C {\mathcal{C}}
\def \D {{D}}

\def \N {{\mathcal{N}}}

\def \N {{\mathcal{N}}}
\def \deq {{\,\overset{{\rm def.}}{=}\,}}

\def \ss {\scriptstyle}

\newcommand{\figref}[1]{{Fig.\ \ref{#1}}}
\newcommand{\secref}[1]{{sec.\ \ref{#1}}}

\DeclareRobustCommand{\SkipTocEntry}[4]{}

\definecolor{colorTC}{rgb}{.2,.7,.2}

\definecolor{amethyst}{rgb}{0.6, 0.4, 0.8}

\definecolor{blue3}{RGB}{31, 119, 180}
\definecolor{red3}{RGB}{	214, 39, 40}
\definecolor{orange3}{RGB}{255, 127, 14}
\definecolor{green3}{RGB}{44, 160, 44}

\begin{document}

\begin{titlepage}
\setcounter{page}{1} \baselineskip=15.5pt
\thispagestyle{empty}
$\quad$
\vskip 70 pt

\begin{center}
{\fontsize{18}{18} \bf Regulating Loops in de Sitter Spacetime}
\end{center}

\vskip 20pt
\begin{center}
\noindent
{\fontsize{12}{18}\selectfont  Akhil Premkumar}
\end{center}


\begin{center}
\vskip 4pt
\textit{ {\small Department of Physics, University of California at San Diego,  La Jolla, CA 92093, USA}}

\end{center}

\vspace{0.4cm}
 \begin{center}{\bf Abstract}
 \end{center}
\noindent Perturbative quantum field theory (QFT) calculations in de Sitter space are riddled with contributions that diverge over time. These contributions often arise from loop integrals, which are notoriously hard to compute in de Sitter. We discuss an approach to evaluate loop integrals that contribute to equal-time correlators of a scalar field theory in a fixed de Sitter background. Our method is based on the Mellin-Barnes representation of correlation functions, which allows us to regulate loop divergences by adjusting the masses of the fields, or by gently deforming the underlying de Sitter spacetime. The resulting expressions have a similar structure as a standard answer from dimensional regularization in flat space QFT. These features of the regulator are illustrated with two examples, worked out in detail. Along the way, we illuminate the physical origin of these divergences and their interpretation with the machinery of the dynamical renormalization group. Our approach regulates the IR divergences of massless and massive particles in the same way. For massless scalars, the loop corrections can be incorporated as systematic improvements to the stochastic inflation framework, allowing for a more precise description of the IR dynamics of such fields in de Sitter.


\end{titlepage}

\setcounter{page}{2}

\restoregeometry

\begin{spacing}{1.2}
\newpage
\setcounter{tocdepth}{3}
\tableofcontents
\end{spacing}

\setstretch{1.1}
\newpage

\section{Introduction}


Inflationary theory \cite{Guth:1980zm,Linde:1981mu,Starobinsky:1980te} posits that the early universe underwent a period of approximately de Sitter (dS) expansion. During this period, inflaton modes of large comoving wavelength leave the horizon, only to re-enter at a later time and seed the structure we see in the night sky.
Thus, the quantum dynamics of these super-Hubble fluctuations become ingrained in our cosmological observations. In particular, the equal time in-in correlation functions of light scalar fields encode a great deal of information about the inflationary era that could be revealed by measurements of primordial non-Gaussianity \cite{Meerburg:2019qqi}. However, not enough is known about these correlators beyond tree level. This situation becomes untenable as our measurements improve in precision, especially in light of the fact that loop calculations lead to infrared (IR) divergences and unbounded time-dependent `secular' growth \cite{Ford:1984hs,Antoniadis:1985pj,Tsamis:1994ca, Tsamis:1996qm,Tsamis:1997za,Burgess:2010dd,Marolf:2010zp, Marolf:2011sh,Marolf:2012kh,Beneke:2012kn,Akhmedov:2013vka,Anninos:2014lwa, Akhmedov:2017ooy,Hu:2018nxy,Akhmedov:2019cfd,Weinberg:2005vy,Weinberg:2006ac} (see appx.\ \ref{sec:AnomGeometric}). Such secular terms appear at all orders of perturbation theory and will be our primary concern in this paper.



In this work, we develop a method to compute loop diagrams that contribute to equal time correlation functions of a scalar field theory on a fixed de Sitter background.
Even simple 1-loop diagrams of this sort are difficult to calculate \cite{Green:2020txs,Marolf:2010zp}. The basic problem is the lack of time translational invariance in dS, which means time appears explicitly in the momentum integrals. Such integrals are not scaleless, hampering our ability to compute them with the usual bag of tricks we employ in flat space.
A solution to this problem was recently introduced in the form of Soft de Sitter Effective Theory (SdSET) \cite{Cohen:2020php,Cohen:2021fzf} where, in the long wavelength limit, the time integrals factorize and separate from integrals over 3-momenta. This returns the scalelessness of the momentum integrals, allowing us to tame the divergences without violating the symmetries of the underlying dS spacetime. However, we still need to match the EFT with the UV theory, and it would be desirable to have a way of regulating loop integrals on the UV side that shares all the nice properties of the regulator we use in the EFT ({\it dynamical} dimensional regularization). Such a procedure must also be generalizable so that we don't have to invent a new way of doing the integral for every diagram we encounter.

Our method starts with the following simple observation: dS spacetime has dilatation invariance in place of the time translation invariance of flat space. The latter allows us to Fourier transform the time variable, which suggests that the transform best suited for dS should have, as its basis, the eigenfunctions of the dilatation generator. Mellin transforms have exactly that property
\cite{Sleight:2019mgd,Sleight:2019hfp,Sleight:2021plv}. Once we switch to Mellin space, the momentum and time integrals decouple, and the divergences manifest as overlaps of certain poles of the integrand. These overlaps may be removed by introducing tiny shifts to these poles, in much the same way that loops in flat space QFT are regulated by tweaking the number of dimensions \cite{Bollini:1972ui,tHooft:1973mfk}. In fact, the connection was first made in \cite{Boos:1990rg}, and Mellin-Barnes (MB) integrals have been used widely in evaluating sophisticated Feynman diagrams in particle physics \cite{Smirnov:2009up, Tausk:1999vh, Anastasiou:2005cb, Czakon:2005rk}. We employ some of the tools developed in these papers in the present work, thereby placing our method on a well-established foundation of computational techniques.

The final expressions of our loop calculations bear a strong resemblance to a standard dimreg answer, with the secular growth encoded in diverging $\Gamma$ functions. Once isolated, such divergences can be interpreted with the machinery of the {\it dynamical} renormalization group (DRG) \cite{Burgess_2010,Green:2020txs,Chen:1995ena,Boyanovsky:2004gq,Podolsky:2008qq,McDonald:2006hf}. This procedure resums large secular logs in the same way regular RG operates on UV logs in flat space. In the effective theory language, the DRG evolution of composite operators of a massless scalar naturally leads to the framework of Stochastic Inflation, which describes the probability distribution of the scalar field as a function of time \cite{Starobinsky:1986fx,Nambu:1987ef,Starobinsky:1994bd,Cohen:2020php,Baumgart:2019clc,Gorbenko:2019rza,Mirbabayi:2019qtx}. Stochastic Inflation is the conceptual basis for slow-roll eternal inflation, and even subtle changes to the time dependence in this picture can have a profound impact on the phase transition to eternal inflation. Loop effects introduce such corrections to this framework at higher order, as demonstrated with SdSET \cite{Cohen:2021fzf}. Our method allows us to draw the same conclusion, by performing loop calculations in the full theory itself.



The paper is structured as follows: In \secref{sec:dSLoopInMellin} we outline the transition to Mellin space and explain how divergences are encoded in this representation.
Next, we review the MB representation of a 4-pt function in \secref{sec:4ptFn}, and use it to compute the $O(\lambda)$ contribution to $\langle \phi^2 \phi^2 \rangle$ and $\langle \phi^3 \phi \rangle$ in \secref{sec:AnomDim} and \ref{sec:phi3phiMixing}. We explain the calculations in detail, to be pedagogical, but most of the intermediate steps are mechanical and can be automated \cite{Czakon:2005rk,Anastasiou:2005cb}. We also identify the issue of requiring more than one parameter to regulate certain integrals and offer some suggestions to resolve this.
By way of a quantitative example, we have calculated the anomalous dimension of the $\phi^2$ operator over a range of masses and summarized them in table.\ \ref{tbl:AnomDimensionsVsMasses}.
We conclude in \secref{sec:Conclusion} by reviewing the lessons learned from our calculations and contemplating future directions.


\section{dS loops in Mellin space}
\label{sec:dSLoopInMellin}

Perturbative QFT calculations in dS are plagued by a variety of divergences. One particular kind, the secular growth terms, causes the naive perturbation expansion to break down at late times. Such contributions are often furnished by loop integrals, which are difficult to compute in dS.
The problem becomes more tractable if we represent the correlation functions in Mellin space.

\subsection{General structure}


Consider a scalar field $\phi$ with mass $m$ in a fixed dS background with the metric $\d s^2 = a(\tau)^2 (-\d\tau^2 + \d \x^2)$, where $\d \x$ is a line element in $\D$ space dimensions, $\tau$ is the conformal time and $a(\tau) = -1/H \tau$. For a free scalar field one expands the field in modes according to
\begin{equation}
  \phi(\x,\tau) = \int \frac{\d^{\D} k}{(2 \pi)^{\D}} e^{i \k \cdot \x} \
  \{ v_{\k}(\tau)a_{\k}+ v^*_{\k}(\tau)a^{\dagger}_{\k} \} \ .
\end{equation}
In the Bunch-Davies vacuum the modes are given by
\begin{gather}
  v_\k(\tau) = e^{-\frac{i\pi}{4}} e^{-\frac{\pi \nu}{2}} \frac{\sqrt{\pi}}{2} H^{\frac{\D-1}{2}} (-\tau)^\frac{\D}{2} {\rm H}^{(1)}_{i \nu}(-k \tau) \ , \\[0.75em]
  \nu \deq i \sqrt{\frac{\D^{2}}{4}-\frac{m^{2}}{H^{2}}} . \label{eq:nuDefn}
\end{gather}
%
where ${\rm H}^{(1)}_{i \nu}$ is the Hankel function of the first kind (we follow the convention in \cite{Sleight:2019mgd}). We are interested in calculating correlation functions of the field at a fixed (late) time using the in-in/Schwinger-Keldysh formalism (see \cite{Weinberg:2005vy,Chen:2017ryl} for a review). Such calculations involve integrals of the schematic form
\begin{align}
  \langle \phi_{\k_1}^{\nu}(\tau_0) \cdots &\phi_{\k_n}^{\nu}(\tau_0) \rangle_{\rm in-in}
  \nonumber \\[0.75em]
  &\supset
  \prod_i \int^{\tau_0} \d \tau_i a(\tau_i)^{D+1}
  \prod_j \int \frac{\d^D p_j}{(2 \pi)^D}
  \dots {\rm H}_{i \nu}(a_m(\k,\p) \tau_i) \,
    {\rm H}_{i \nu}(a_{m+1}(\k,\p) \tau_{i+1}) \dots
  \label{eq:dSCorrStructure}
\end{align}
%
where $a_m(\k, \p)$ are some linear combinations of the 3-momenta. The $\supset$ symbol indicates that the correlator is a sum of many terms like the one on the r.h.s. The evaluation of this integral is made difficult by the fact that the variables of integration, $\tau_i$ and $\p_j$, are trapped as arguments of Hankel functions.
There are few special values of $\nu$ for which ${\rm H}_{i\nu}(-k \tau)$ has a simpler functional form, but even in those cases calculations involving loop integrals are cumbersome\footnote{For instance, the two point function of a conformal mass scalar has the form $\langle \phi \phi \rangle \sim e^{i k (\tau-\tau')}/k$, where the exponential makes it difficult to evaluate the loop integral with techniques we use in flat space calculations (see \cite{Green:2020txs} for more details). Life can be made simpler by imposing a hard cutoff, but this could generate unphysical logs.}. To make progress, we rely on the following convenient MB representations \cite{Watson1995treatise},
%
\begin{equation}
  \begin{aligned}
    i \pi e^{-\frac{\pi \nu}{2}} {\rm H}^{(1)}_{i \nu}(z)
      &= \int_{c-i\infty}^{c+i\infty} \frac{\d s}{2 \pi i} \,
          \Gamma\left( s + \frac{i \nu}{2} \right)
          \Gamma\left( s - \frac{i \nu}{2} \right)
          \left( -\frac{i z}{2} \right)^{-2s} \\[1em]
    -i \pi e^{\frac{\pi \nu}{2}} {\rm H}^{(2)}_{i \nu}(z)
      &= \int_{c-i\infty}^{c+i\infty} \frac{\d s}{2 \pi i} \,
          \Gamma\left( s + \frac{i \nu}{2} \right)
          \Gamma\left( s - \frac{i \nu}{2} \right)
          \left( \frac{i z}{2} \right)^{-2s}
  \end{aligned}
  \label{eq:MBreprHankel}
\end{equation}
where $c > |\nu|/2$. These representations have been used to study late-time tree level correlation functions in \cite{Sleight:2019mgd,Sleight:2019hfp}. Building on that work, we will explore whether the same approach is fruitful in analyzing loop integrals in dS. To outline the procedure we begin by substituting \eqref{eq:MBreprHankel} into \eqref{eq:dSCorrStructure} and changing the order of integration,
\begin{align}
  \langle \phi_{\k_1}^{\nu}(\tau_0) \cdots \phi_{\k_n}^{\nu}(\tau_0) \rangle_{\rm in-in}
  \supset
  \prod_\ell \int &\d s_\ell  \,
    \Gamma\left( s_\ell + {\ss \frac{i \nu}{2}} \right)
    \Gamma\left( s_\ell - {\ss \frac{i \nu}{2}} \right)
  \prod_i \int^{\tau_0} \d \tau_i a(\tau_i)^{D+1} (-\tau_i)^{-2s_\ell} \nonumber \\
  &\times
  \prod_j \int \frac{\d^D p_j}{(2 \pi)^D} \dots a_m(\k, \p)^{-2 s_\ell} \, a_{m+1}(\k,\p)^{-2 s_{\ell'}} \dots .
\end{align}
We have `released' the variables $\tau_i$ and $\p_j$, at the cost of introducing an MB integral for each Hankel function. The Mellin variables $s_\ell$ label the eigenstates of the dilatation generator \cite{Sleight:2021plv}. In the late-time limit, $\tau_0 \to 0$, the time integrals reduce to $s$-conserving delta functions at each vertex \cite{Sleight:2019mgd,Sleight:2019hfp}; dilatation invariance of de Sitter space leads to conservation of $s$ the same way that translation invariance implies conservation of 3-momenta $\k$. We are then left with
\begin{align}
  \langle \phi_{\k_1}^{\nu}(\tau_0) \cdots \phi_{\k_n}^{\nu}(\tau_0) \rangle_{\rm in-in}
  \supset
  \prod_\ell \int &\d s_\ell  \,
    \Gamma\left( s_\ell + {\ss \frac{i \nu}{2}} \right)
    \Gamma\left( s_\ell - {\ss \frac{i \nu}{2}} \right)
    Q({\bm s}) \delta(s_\ell + \dots) \delta(s_{\ell''} + \dots) \dots
  \nonumber \\
  &\times
  \prod_j \int \frac{\d^D p_j}{(2 \pi)^D} \dots a_m(\k, \p)^{-2 s_\ell} \, a_{m+1}(\k,\p)^{-2 s_{\ell'}} \dots .
\end{align}
where $Q({\bm s}) \equiv Q(s_1, \dots, s_\ell, \dots)$ is some ratio of polynomials of the Mellin variables. The form of the r.h.s.\ reflects the fact that dilatation and translation do not commute. The momentum integrals are now manifestly scaleless. Evaluating these introduces new $s_\ell$-dependent $\Gamma$ functions,
\begin{equation}
  \langle \phi_{\k_1}^{\nu}(\tau_0) \cdots \phi_{\k_n}^{\nu}(\tau_0) \rangle_{\rm in-in}
  \supset
  {\prod_\ell}^\prime \int \d s_\ell  \,
  \Gamma\left( s_\ell + {\ss \frac{i \nu}{2}} \right)
  \Gamma\left( s_\ell - {\ss \frac{i \nu}{2}} \right)
  Q({\bm s})
  \frac{
    \Gamma\left( D - s_{\ell'} - \dots \right) \dots
  }{
    \Gamma\left( s_{\ell} - \dots \right) \dots
  }
  b_{\ell}(\k)^{s_\ell}
  \label{eq:dSCorrFinalForm}
\end{equation}
where $b_\ell(\k)$ are ratios involving the external momenta, and the prime on ${\prod_j}^\prime$ indicates that we have applied the delta functions in $s$. All that remains is the evaluation of a multidimensional MB integral \eqref{eq:dSCorrFinalForm}. Integrals of this form have been studied extensively as a tool to simplify certain Feynman diagrams in particle physics \cite{Smirnov:2004ym, Smirnov:2009up, Tausk:1999vh, Anastasiou:2005cb, Czakon:2005rk}. Drawing on that literature, we turn our attention to extracting and understanding the divergences contained in \eqref{eq:dSCorrFinalForm}.

\subsection{Divergences in MB integrals}
\label{sec:DivsInMB}

Consider the simple case of an MB integral
\begin{equation}
  K(x) =
   \int_{-i \infty}^{+i \infty} \frac{\d s}{2 \pi i}
     \Gamma(\textcolor{red}{s-a}) \Gamma(\textcolor{blue}{-s}) x^{-s} .
  \label{eq:MBtoy1}
\end{equation}
The poles of the integrand are the poles of the $\Gamma$ functions. These are at $s_\star = a, a-1, a-2, a-3, \dots$ which are the `\textcolor{red}{left}' poles and at $s_\star = 0,1,2,3, \dots$ which are the `\textcolor{blue}{right}' poles. The contour is a straight line that runs parallel to the $\Im(s)$ axis and it must separate all left poles from all right poles. This is the Mellin contour prescription. We may then close the contour on either the left or right half plane, picking up the residues at the poles.

\begin{figure}
  \begin{subfigure}{\textwidth}
  \centering
    \begin{tikzpicture}[scale=1]
      \draw[thick,->] (-5,0) -- (5,0) node[anchor=south] {$\Re(s)$};
      \draw[thick,->] (0,-1.5) -- (0,1.5) node[anchor=west] {$\Im(s)$};

      \foreach \z in {0,1,2}
        \filldraw[draw=black,fill=blue] ({2*\z},0) circle (4pt) node[anchor=north west,yshift=-3mm] {$\textcolor{blue}{\scriptstyle \z}$};
      \foreach \z in {0,1,2}
        \filldraw[draw=black,fill=red] ({-0.05-2*\z},0) circle (4pt) node[anchor=south east,yshift=3mm] {$\textcolor{red}{\scriptstyle \ifthenelse{\z=0}{}{-} \z}$};


      \node at (-3,1.5) {\underline{With $a = 0$}};
    \end{tikzpicture}
  \caption{\label{fig:PoleOverlap} Poles of $\Gamma(\textcolor{red}{s}) \Gamma(\textcolor{blue}{-s})$ }
  \end{subfigure}
  \\[1em]
  \begin{subfigure}{\textwidth}
  \centering
    \begin{tikzpicture}[scale=1]
      \draw[thick,->] (-5,0) -- (5,0) node[anchor=south] {$\Re(s)$};
      \draw[thick,->] (0,-1.5) -- (0,1.5) node[anchor=west] {$\Im(s)$};

      \foreach \z in {0,1,2}
        \filldraw[draw=black,fill=blue] ({2*\z},0) circle (4pt) node[anchor=north west,yshift=-3mm] {$\textcolor{blue}{\scriptstyle \z}$};
      \foreach \z in {0,1,2}
        \filldraw[draw=black,fill=red] ({-0.5-2*\z},0) circle (4pt) node[anchor=south,xshift=1mm,yshift=3mm] {$\textcolor{red}{\scriptstyle \ifthenelse{\z=0}{}{-\z - \e}}$};
      \node[anchor=south east,yshift=3mm] at (-0.5,0) {$\textcolor{red}{\scriptstyle -\e}$};

      \draw[thick,->,draw=magenta] (-.25,-1.5) -- (-.25,1.5);
      \node at (-0.5,-1) {$\C$};
      \node at (-3,1.5) {\underline{With $a=-\e < 0$}};

      \draw[thin,->] (1,-0.75) node[below] {$\sim \Gamma(\e)$} to [out=120,in=-30] (0.2,-0.2);
      \draw[thin,->] (3,1) node[above] {\small finite} to [out=200,in=90] (2,0.2);
      \draw[thin,->] (3,1) to [out=-20,in=90] (4,0.2);
    \end{tikzpicture}
  \caption{\label{fig:PoleOverlapFixed} Poles of $\Gamma(\textcolor{red}{s+\e}) \Gamma(\textcolor{blue}{-s})$ }
  \end{subfigure}
  \\[1em]
  \caption{Regulating divergences in MB integrals. The overlap of the zeroth left and right poles in (a) makes \eqref{eq:MBtoy1} undefined because no contour can separate all the left poles from the right ones. In (b) we have shifted the left poles to the left by $\e$, allowing us to drive a contour between the two kinds of poles. In the limit $\e \to 0$ we approach the situation in (a), known as {\it contour pinching}, which results in a divergence $\Gamma(\e)$. This is simply the residue at $s_\star=0$, as we close the contour on the right hand plane. The residues at the other poles are finite in the same limit.}
\end{figure}
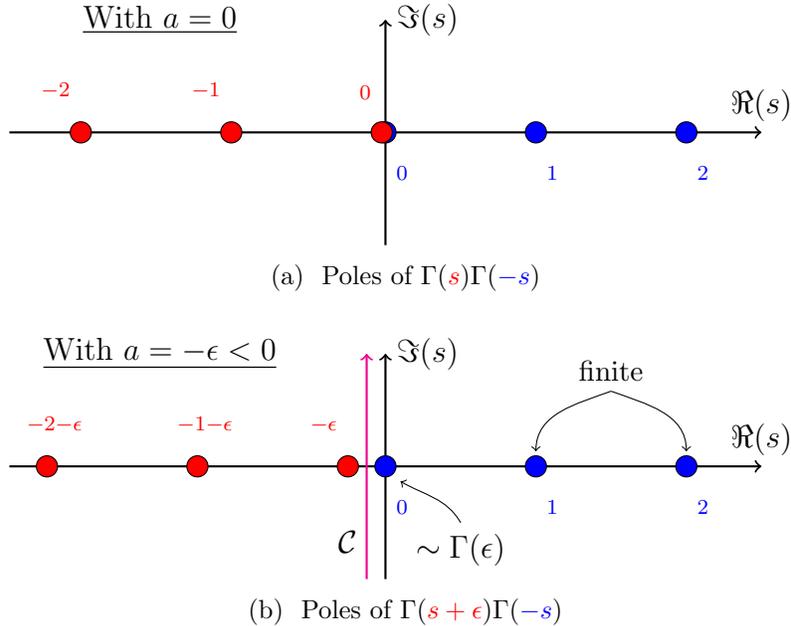

Now consider the case where $a=0$. This leads to an overlap of the zeroth left and right poles as shown in \figref{fig:PoleOverlap}. In this situation no choice of contour can separate all left poles from the right ones. Instead, if we set $a = -\e$, where $\e$ is a vanishingly small positive number, we have the situation shown in \figref{fig:PoleOverlapFixed}. The overlap is removed and a contour $\C$ can be driven between the left and right poles. As $\e$ approaches zero the contour is said to be `pinched'. The value of the integral at $\e=0$ is defined by analytic continuation of the integral at $\e > 0$.

If we close $\C$ on the right hand plane we pick up the residues at the right poles. However, notice that the residue at the zeroth right pole $s_\star = 0$ is
\begin{equation}
  {\rm Res}[\Gamma(s+\e) \Gamma(-s) x^{-s}; s=0] = -\Gamma(\e),
  \label{eq:DivergentResidue}
\end{equation}
which blows up as $\e \to 0$. In other words, an overlap of left/right poles signals divergences in the MB integral. Also note how the divergence appears as a $\Gamma(\e)$, in the same way it does for a flat space loop integral computed with dimreg\footnote{In fact, the Mellin approach produces the same answer in flat space, that we obtain by the conventional methods. See Ch.\ 4 of \cite{Smirnov:2004ym} for an instructive example that uses the MB representation of Feynman propagators to compute the one-loop self-energy graph in QED.}.

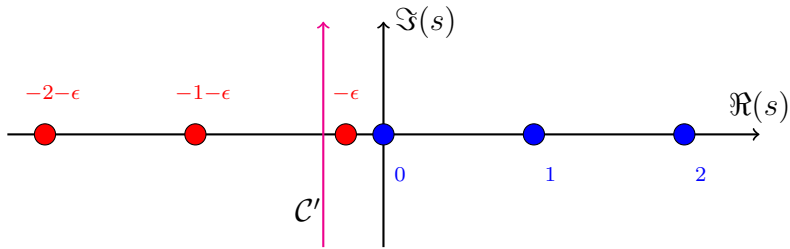
\begin{figure}
\centering
  \begin{tikzpicture}[scale=1]
    \draw[thick,->] (-5,0) -- (5,0) node[anchor=south] {$\Re(s)$};
    \draw[thick,->] (0,-1.5) -- (0,1.5) node[anchor=west] {$\Im(s)$};

    \foreach \z in {0,1,2}
      \filldraw[draw=black,fill=blue] ({2*\z},0) circle (4pt) node[anchor=north west,yshift=-3mm] {$\textcolor{blue}{\scriptstyle \z}$};
    \foreach \z in {0,1,2}
      \filldraw[draw=black,fill=red] ({-0.5-2*\z},0) circle (4pt) node[anchor=south,xshift=1mm,yshift=3mm] {$\textcolor{red}{\scriptstyle \ifthenelse{\z=0}{}{-\z - \e}}$};
    \node[anchor=south,yshift=3mm] at (-0.5,0) {$\textcolor{red}{\scriptstyle -\e}$};

    \draw[thick,->,draw=magenta] (-.8,-1.5) -- (-.8,1.5);
    \node at (-1,-1) {$\C^\prime$};

  \end{tikzpicture}
  \caption{\label{fig:IncorrectContour}A contour that does not separate left/right poles.}
\end{figure}

What happens if the contour does not separate the left and right poles? Suppose we chose to integrate over a contour $\C^\prime$ that intersects $\Re(s)$ to the left of the pole at $s_\star = -\e$, as in \figref{fig:IncorrectContour}. Then, this pole must be included in the sum of residues. But that means $K'(x) = {\rm Res}[s=-\e] + {\rm Res}[s=0] + \dots = \Gamma(\e) - \Gamma(\e) + \dots$. That is, there is no divergence, even in the limit $\e \to 0$ when the zeroth poles overlap.
The situations in \figref{fig:PoleOverlapFixed} and \figref{fig:IncorrectContour} differ only by the placement of the pole at $s_\star = -\e$. Therefore we can write
\begin{equation}
\int_{\C} \frac{\d s}{2 \pi i} \Gamma(s+\e) \Gamma(-s) x^{-s}
  =
  +{\rm Res}[s=-\e] +
  \int_{\C'} \frac{\d s}{2 \pi i} \Gamma(s+\e) \Gamma(-s) x^{-s} .
  \label{eq:SeparatingADivergence1}
\end{equation}
The integral over $\C'$ is finite and the divergence manifests in the residue at $-\e$. Incidentally, this also suggests an algorithm to identify and separate the divergence from any Mellin integral. If we keep the contour {\it fixed} in \figref{fig:PoleOverlapFixed} and decrease $\e$, the zeroth left pole will cross over to the right at some point.
We then end up in the same situation as \figref{fig:IncorrectContour}, except that the contour is still $\C$. To recover the original integral we'll need to add the residue at $s_\star = -\e$, just as we did in \eqref{eq:SeparatingADivergence1}, thereby isolating the divergence as a separate term. This is the basis of the procedure, first introduced in \cite{Tausk:1999vh}, to extract divergences from Mellin integrals over several variables.

\subsection{\texorpdfstring{$N$}{N}-dimensional MB integrals}

By iterating the same basic steps discussed above, we can identify and separate divergences in an $N$-dimensional MB integral of the general form (cf.\ \eqref{eq:dSCorrFinalForm})
%
\begin{equation}
  K(x_1, x_2,\dots, x_N) =
    \int_{-i \infty}^{i \infty} \frac{\d s_1}{2 \pi i} \dots
    \int_{-i \infty}^{i \infty} \frac{\d s_N}{2 \pi i} \,
    \frac{
      \prod_i \Gamma \left(U_{i}(\bm{s})\right)
    }{
    \prod_j \Gamma \left(V_{j}(\bm{s})\right)
    } x_{1}^{-s_{1}} \cdots x_{N}^{-s_{N}} ,
    \label{eq:MBgeneralform}
\end{equation}
where $x_n$ are the ratios of kinematic variables, $\bm{s}=\left(s_{1}, \ldots, s_{N}\right)$, and the arguments of the $\Gamma$ functions in the MB integrand are
\begin{equation}
  \begin{aligned}
    U_{i}(\bm{s}) &\deq a_{i} +  \sum_\ell b_{i \ell} s_\ell \\
    V_{j}(\bm{s}) &\deq a'_{i} + \sum_\ell b'_{i \ell} s_\ell
  \end{aligned}
  \label{eq:GammaFnArguments}
\end{equation}
where the constants $a_{i}, b_{i\ell}$ etc.\ are reals. The integral in \eqref{eq:MBgeneralform} is just an extension of \eqref{eq:MBtoy1} to $N$ Mellin variables. The poles of this integral are the poles of the $\Gamma$ functions in the numerator, that is, those $\bm s$ where
\begin{equation}
  U_{i}(\bm{s}) = -n , \qquad n \in \mathbb{Z}_0 . \label{eq:MBPoles}
\end{equation}
As before, an MB integral is well-defined if the contours separate the left and right poles. For a multidimensional MB integral like \eqref{eq:MBgeneralform} this condition is equivalent to the requirement
{
\def \sC {{\Re(s^\C)}}
\def \sC {{s^\C}}
\begin{equation}
  U_i(\C) > 0 \quad \forall \ i \label{eq:MellinContourPrescription}
\end{equation}
where $U_i(\C)$ is the real part of $U_i$ evaluated on the contour $\C$. For example, consider the situation in \figref{fig:MellinContour}. If $U_1(s) = a+s$, then $U_1(\C) = a + \sC > 0$ if we choose a contour which intersects the $\Re(s)$ axis at $\sC > -a$. Similarly, if $U_2(s) = b-s$ then $U_2(\C) = b - \sC > 0 \implies \sC < b$.
Taken together, an integral with both these $\Gamma$ functions requires $-a < \sC < b$, which is exactly the condition that the contour must separate the left/right poles.
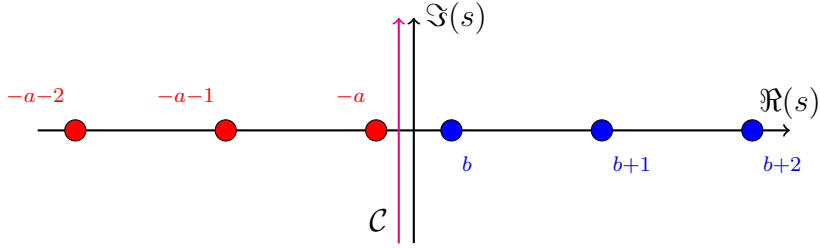
\begin{figure}
  \centering
  \begin{tikzpicture}[scale=1]
    \draw[thick,->] (-5,0) -- (5,0) node[anchor=south] {$\Re(s)$};
    \draw[thick,->] (0,-1.5) -- (0,1.5) node[anchor=west] {$\Im(s)$};

    \foreach \z in {0,1,2}
      \filldraw[draw=black,fill=blue] ({0.5+2*\z},0) circle (4pt) node[anchor=north west,yshift=-2mm] {$\textcolor{blue}{\ss b \ifthenelse{\z=0}{}{+\z}}$};
    \foreach \z in {0,1,2}
      \filldraw[draw=black,fill=red] ({-0.5-2*\z},0) circle (4pt) node[anchor=south east,yshift=2mm] {$\textcolor{red}{\ss -a \ifthenelse{\z=0}{}{-\z}}$};

    \draw[thick,->,draw=magenta] (-.2,-1.5) -- (-.2,1.5);
    \node[left] at (-.2,-1.2) {$\C$};
  \end{tikzpicture}
  \caption{\label{fig:MellinContour} Pole structure of an MB integral $\int_\C \frac{\d s}{2 \pi i} \Gamma(a+s) \Gamma(b-s) x^{-s}$.
  The contour satisfies $a < \sC < b$, which is equivalent to the condition that the arguments of the $\Gamma$ functions in the integrand are positive when evaluated on the contour (cf.\ \eqref{eq:MellinContourPrescription}).}
\end{figure}
}

Generalizing this further, the condition for the $n^{\rm th}$ pole of $\Gamma(U_i)$ to be on the `correct' side is $U_i(\C) + n > 0 $. That is, an $n^{\rm th}$ right pole will be on the right side of the contour and an $n^{\rm th}$ left pole will be on the left if this inequality is satisfied.
Otherwise it means that the pole has crossed the contour. Thus, we have the following pole crossing condition for the $n^{\rm th}$ pole,
\begin{equation}
  U_i(\C) + n < 0 . \label{eq:PoleCrossingCondition}
\end{equation}
The inequalities \eqref{eq:MellinContourPrescription} and \eqref{eq:PoleCrossingCondition} allow us determine where a pole is, even in the $N$-dimensional case when it becomes difficult to visualize the location of the poles and contours.

\subsection{Analytic continuation}
\label{sec:GeneralAC}

\begin{figure}
  \begin{subfigure}{\textwidth}
  \centering
    \begin{tikzpicture}[scale=1]
      \draw[thick,->] (-6,0) -- (6,0) node[anchor=south] {$\Re(s_\ell)$};
      \draw[thick,->] (0,-2) -- (0,2) node[anchor=west] {$\Im(s_\ell)$};

      \begin{scope}[shift={(2.8,0.6)}]
        \foreach \z in {0,1}
          \filldraw[draw=black,fill=blue] ({2*\z},0) circle (4pt);
      \end{scope}

      \begin{scope}[shift={(-0.8,-0.6)}]
        \foreach \z in {0,1,2,3}
          \filldraw[draw=black,fill=blue] ({2*\z},0) circle (4pt);
      \end{scope}

      \begin{scope}[shift={(2.75,0.6)}]
        \foreach \z in {0,1,2,3,4}
          \filldraw[draw=black,fill=red] ({-2*\z},0) circle (4pt);
      \end{scope}

      \begin{scope}[shift={(-0.85,-0.6)}]
        \foreach \z in {0,1,2}
          \filldraw[draw=black,fill=red] ({-2*\z},0) circle (4pt);
      \end{scope}

      \draw[thin,->] (-1.1,-1.5) node[left] {overlap} to [out=0,in=-90] (-0.825,-1);
      \draw[thin,->] (3,1.5) node[right] {overlap} to [out=180,in=90] (2.775,1);
    \end{tikzpicture}
  \caption{\label{fig:PolesUnparameterized} The original integrand, with all $\e_k = 0$, has overlapping poles.}
  \end{subfigure}
  \\[1em]
  \begin{subfigure}{\textwidth}
  \centering
    \begin{tikzpicture}[scale=1]
      \draw[thick,->] (-6,0) -- (6,0) node[anchor=south] {$\Re(s_\ell)$};
      \draw[thick,->] (0,-2) -- (0,2) node[anchor=west] {$\Im(s_\ell)$};

      \begin{scope}[shift={(0.5,0.6)}]
        \foreach \z in {0,1,2}
          \filldraw[draw=black,fill=blue] ({2*\z},0) circle (4pt);
      \end{scope}

      \begin{scope}[shift={(1,-0.6)}]
        \foreach \z in {0,1,2}
          \filldraw[draw=black,fill=blue] ({2*\z},0) circle (4pt);
      \end{scope}

      \begin{scope}[shift={(-1,0.6)}]
        \foreach \z in {0,1,2}
          \filldraw[draw=black,fill=red] ({-2*\z},0) circle (4pt);
      \end{scope}

      \begin{scope}[shift={(-1.5,-0.6)}]
        \foreach \z in {0,1,2}
          \filldraw[draw=black,fill=red] ({-2*\z},0) circle (4pt);
      \end{scope}

      \draw[thick,->,draw=magenta] (-.3,-2) -- (-.3,2);
      \node at (-0.55,1.7) {$\C$};


    \end{tikzpicture}
  \caption{
    \label{fig:PolesSeparated} The poles are separated initially by setting $\e_k = \e_k^{(0)}$.
  }
  \end{subfigure}
  \\[1em]
  \begin{subfigure}{\textwidth}
  \centering
    \begin{tikzpicture}[scale=1]
      \draw[thick,->] (-6,0) -- (6,0) node[anchor=south] {$\Re(s_\ell)$};
      \draw[thick,->] (0,-2) -- (0,2) node[anchor=west] {$\Im(s_\ell)$};

      \begin{scope}[shift={(2.8,0.6)}]
        \foreach \z in {0,1}
          \filldraw[draw=black,fill=blue] ({2*\z},0) circle (4pt);
      \end{scope}

      \begin{scope}[shift={(-0.8,-0.6)}]
        \foreach \z in {0,1,2,3}
          \filldraw[draw=black,fill=blue] ({2*\z},0) circle (4pt);
      \end{scope}

      \begin{scope}[shift={(2.4,0.6)}]
        \foreach \z in {0,1,2,3,4}
          \filldraw[draw=black,fill=red] ({-2*\z},0) circle (4pt);
      \end{scope}

      \begin{scope}[shift={(-1.2,-0.6)}]
        \foreach \z in {0,1,2}
          \filldraw[draw=black,fill=red] ({-2*\z},0) circle (4pt);
      \end{scope}

      \draw[thick,->,draw=magenta] (-.25,-2) -- (-.25,2);
      \node at (-0.55,1.7) {$\C$};

      \draw[thick,dashed] (2.6,0.6) ellipse (0.5 cm and 0.25 cm);
      \draw[thick,dashed] (-1,-0.6) ellipse (0.5 cm and 0.25 cm);


      \draw[thin,->] (-1.5,-1.5) node[left] {$\sim \Gamma({\bm \e})$} to [out=0,in=-90] (-1,-1);
      \draw[thin,->] (3,1.5) node[right] {$\sim \Gamma({\bm \e})$} to [out=180,in=90] (2.6,1);
    \end{tikzpicture}
  \caption{\label{fig:PolesCrossed} Some poles end up on the wrong side of the contour and nearly overlap when $\e_k \sim 0$}
  \end{subfigure}
  %
  %
  \caption{An MB integral is defined by shifting its poles around till the left/right poles are separated by a straight line contour. Analytic continuation involves reverting these shifts and allowing the poles to cross back to their original position. As a pole crosses the contour, we isolate the contribution at that pole (cf.\ \eqref{eq:SeparatingADivergence1}). Some of the left/right poles (encircled) nearly overlap, resulting in divergences. For an $N$-fold MB integral all this happens across $N$ complex $s_\ell$-planes simultaneously. Note: The poles in these figures have been staggered vertically for clarity; they are all real-valued in our examples. See \secref{sec:GeneralAC} for further explanation.}
\end{figure}
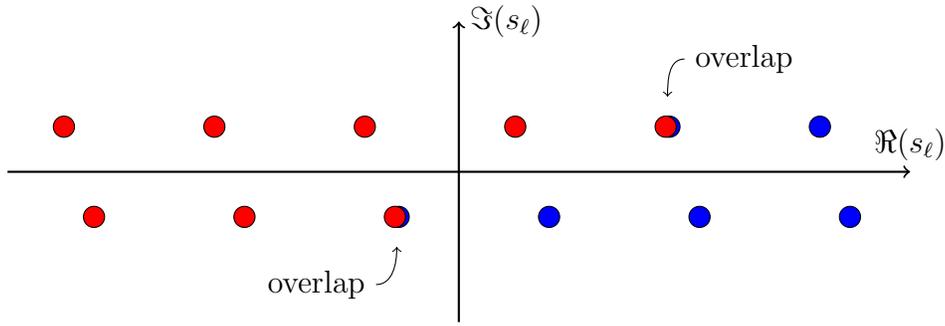
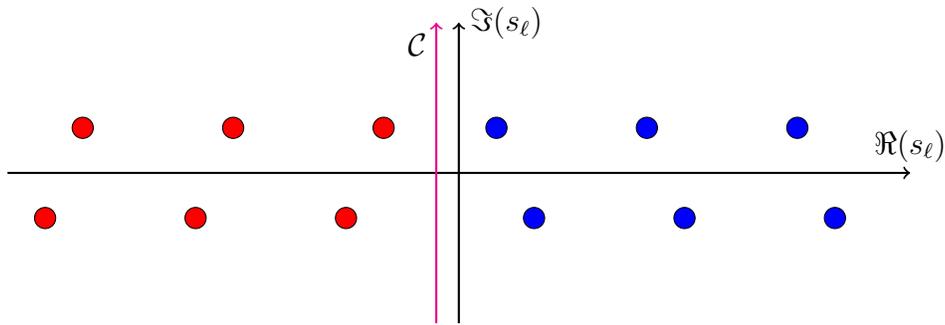
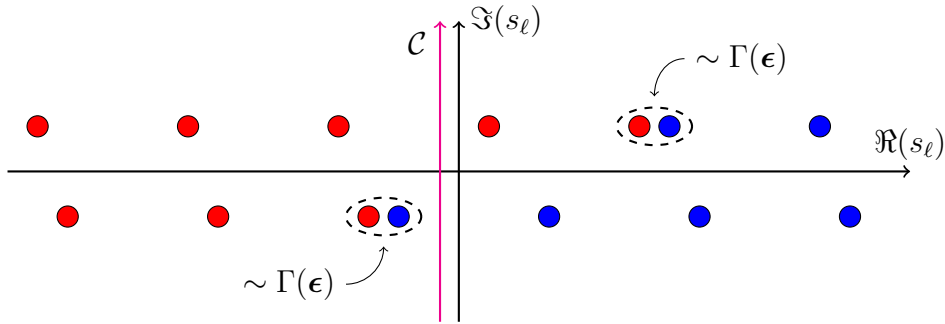

The MB integral in \eqref{eq:MBgeneralform} will have divergences if the integrand has overlapping left/right poles. This makes it impossible to choose a set of straight line contours that satisfy \eqref{eq:MellinContourPrescription} for all $U_i$ (\figref{fig:PolesUnparameterized}). We circumvent this issue by introducing parameters $\e_k$ into our integral that shifts the poles around until we can meet the conditions \eqref{eq:MellinContourPrescription}. That is, we change \eqref{eq:GammaFnArguments} to
\begin{equation}
  \begin{aligned}
    U_{i}(\bm{s}) &\to a_{i} + \sum_\ell b_{i \ell} s_\ell + \sum_k c_{i k} \e_k \\
    V_{j}(\bm{s}) &\to a'_{i} +  \sum_\ell b'_{i \ell} s_\ell + \sum_k c'_{i k} \e_k ,
  \end{aligned}
  \label{eq:GammaFnArgumentsModified}
\end{equation}
and choose the initial values $\e_k = \e_k^{(0)}$ to make the left/right poles separable with straight line contours (\figref{fig:PolesSeparated}). The original integral, for which $\e_k=0$, is defined by analytic continuation of the integral with the modified arguments \eqref{eq:GammaFnArgumentsModified}.
As we take $\e_k \to 0$, some of the poles will cross over to the `wrong' side of the contours (\figref{fig:PolesCrossed}), and we separate the residues at those poles as in \eqref{eq:SeparatingADivergence1}. Around $\e_k \sim 0$ some left/right poles nearly overlap, with the small non-zero values of $\e_k$ keeping them from complete coalescence. The residues at these poles isolate the divergences in the original integral.
We can understand these ideas with a rudimentary example. Consider the double integral
\begin{equation}
  K
  = \int_{\C_z} \frac{\d z}{2 \pi i} \int_{\C_w} \frac{\d w}{2 \pi i} \,
    \Gamma_1(z,w,\e) \Gamma_2(z,w,\e) \dots \Gamma_m(z,w,\e) , \label{eq:RudyExample}
\end{equation}
where $\Gamma_i(z,w,\e) \equiv \Gamma(U_i(z,w,\e))$ are Gamma functions with arguments of the form \eqref{eq:GammaFnArgumentsModified}, and $\C_z$ and $\C_w$ are straight line contours which we close in the right half plane.
These contours separate the left/right poles of the integrand at $\e=\e_0$. Suppose that, as we decrease $\e \to 0$, the first left pole $z_\star$ of $\Gamma_1$ has crossed to the right of $\C_z$ at $\e = \e_1$. Following the discussion around \eqref{eq:SeparatingADivergence1}, we may write
%
\begin{align}
  K &=
  {\rm Res}[\Gamma_1; z_\star] \int_{\C_w} \frac{\d w}{2 \pi i} \Gamma_2(z_\star,w,\e_1) \dots +
  \int_{\C_z} \frac{\d z}{2 \pi i} \int_{\C_w} \frac{\d w}{2 \pi i} \Gamma_1^*(z,w,\e_1) \Gamma_2(z,w,\e_1) \dots \nonumber \\
  &\deq K_1 + K^* .
\end{align}
The asterisk on $\Gamma_1^*$ indicates that the $n=0$ pole of that function is on the `wrong' side. Therefore, the double integral $K^\star$ includes a contribution $-{\rm Res}[\Gamma_1; z_\star]$ (the minus sign is due to the clockwise direction of the contour), which needs to be compensated with a $+{\rm Res}[\Gamma_1; z_\star]$ to recover the original integral $K$. This is exactly the term $K_1$, which is the residue of $K$ at the first pole of $\Gamma_1$.
If $z_\star$ was a right pole that crossed to the left we would have written $\int_{\C_z} \frac{\d z}{2 \pi i} \Gamma_1(z,w,\e_1) = -{\rm Res}[\Gamma_1; z_\star] + \int_{\C_z} \frac{\d z}{2 \pi i} \Gamma_1^*(z,w,\e_1)$.
Next, we concentrate on the $K_1$ term\footnote{Note that $\Gamma_2(z_\star,w,\e_1)$ can have a different $w$ dependence than $\Gamma_2(z,w,\e_1)$.}. As we continue decreasing $\epsilon$, suppose a pole $w_\star$ of $\Gamma_2$ has crossed over at $\e=\e_2$. Then,
\begin{align}
  K_1 &=
  {\rm Res}[\Gamma_1; z_\star] {\rm Res}[\Gamma_2(z_\star) \dots \Gamma_m(z_\star); w_\star]
  + {\rm Res}[\Gamma_1; z_\star] \int_{\C_w} \frac{\d w}{2 \pi i} \Gamma_2^\star(z_\star,w,\e_2) \dots \nonumber \\
  &\deq K_{12} + K_1^* .
\end{align}
%
Similarly, if a pole of $\Gamma_2$ crosses over in $K^\star$ as we decrease $\e$ from $\e_1$, we write $K^* = K_2^* + K^{**}$. We will assume that no more poles cross over as we continue down to $\e \sim 0$. Therefore we can expand the integrals in $K_1^*, K_2^*$ and $K^{**}$ as a Taylor series in $\e$.
It is clear from context which poles are on the wrong side, so we will drop the asterisks on the $K$'s henceforth\footnote{The naming convention for the r.h.s.\ of \eqref{eq:RudyBreakup} is from \cite{Tausk:1999vh}.}. Collecting everything together, we obtain
\begin{equation}
  K \to K_{12} + K_1 + K_2 + K. \label{eq:RudyBreakup}
\end{equation}
This is just a 2-fold version of \eqref{eq:SeparatingADivergence1}. The break up of $K$ depends on the choice of contours, but the final answer will be the same when everything is added up.
Some of the terms in \eqref{eq:RudyBreakup} will contain $\Gamma(\e)$'s, from taking residues at nearly overlapping poles (see \figref{fig:PolesCrossed}). These are the divergences we are looking for. Since \eqref{eq:RudyExample} is a 2-fold integral there can be a maximum of two simultaneous pinches, and the leading behavior around $\e \sim 0$ is at most
\begin{equation}
  K \sim \frac{{\rm const.}}{\e^2} + \frac{{\rm const.'}}{\e} + O(\e^0) .
\end{equation}
The extension to the $N$-dimensional integral \eqref{eq:MBgeneralform} is straightforward. The procedure described above has been developed into an algorithm in \cite{Smirnov:2009up,Anastasiou:2005cb,Czakon:2005rk}. We will illustrate the steps in the examples below.


\section{The tree level 4-point function}
\label{sec:4ptFn}

\begin{figure}
  \centering
  \begin{tikzpicture}

      \draw[thick] (-0.5,4) -- (4.5,4) node[right] {$\tau_0$};
      \draw[thick] (0,4) -- (2,2) -- (4,4);
      \draw[thick] (1.5,4) -- (2,2) -- (2.5,4);
      \filldraw (2,2) circle (2pt) node[below] {$\tau$};

      \filldraw (0,4) circle (1pt) node[above] {$\k_1$};
      \filldraw (1.5,4) circle (1pt) node[above] {$\k_2$};
      \filldraw (2.5,4) circle (1pt);
      \node[above] at (2.7,4) {$\k_3$};
      \filldraw (4,4) circle (1pt) node[above] {$\k_4$};

      \draw[->,thick] (-1,2.25) -- (-1,3.5);
      \node[below] at (-1,2.25) {time};
  \end{tikzpicture}
  \caption{\label{fig:4ptcorrtree} The 4-pt correlation function at tree level.}
\end{figure}
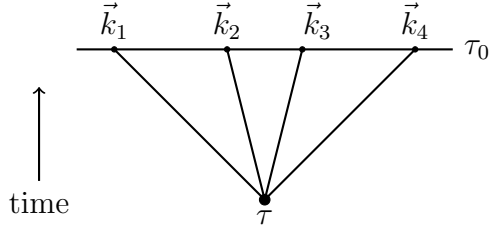

The calculations in this paper begin with the contraction of the tree level 4-point function of scalar fields shown in \figref{fig:4ptcorrtree} (we are working with a $\lambda \phi^4$ interaction). The MB representation of such functions is detailed in \cite{Sleight:2019mgd}. We will summarize just the basic elements required to set up our loop integrals, using the same conventions as that paper.
First, the Mellin-Barnes representation for the dS bulk-to-boundary propagator is
\begin{equation} \label{eq:MellinB2b}
  F_{\pm, \k}^{(\nu)}\left(\tau ; \tau_{0}\right)
  = (-\tau)^{\frac{\D}{2}-i \nu} \mathcal{N}_{\nu}\left(\tau_{0}\right)
    \int_{-i \infty}^{i \infty} \frac{\d s}{2 \pi i} e^{\delta_{\nu}^{\pm}(s)}
    \Gamma \left(s+\frac{i \nu}{2}\right) \Gamma \left(s-\frac{i \nu}{2}\right) \left(-\frac{\tau k}{2}\right)^{-2 s+i \nu}
\end{equation}
where $\tau$ is a point in the bulk and the late time $\tau_0 \to 0$. The contour is a vertical line that intersects the real axis to the right of the pole at $s_\star=-\frac{i\nu}{2}$. The other symbols are
\begin{align}
  \delta_{\nu}^{\pm}(s) &\deq \mp i \pi\left(s+\frac{i \nu}{2}\right) \\
  \mathcal{N}_{\nu}\left(\tau_{0}\right) &\deq \left(-\tau_{0}\right)^{\frac{\D}{2}+i \nu} \frac{\Gamma(-i \nu) H^{\D-1}}{4 \pi}
\end{align}
The $+(-)$ sub-indices indicate the contributions from the (anti)-time-ordered branches of the in-in contour. The four point correlator of \figref{fig:4ptcorrtree} is given by
\begin{equation} \label{eq:Mellin4pt1}
  \left\langle \phi_{\vec{k}_{1}}^{\left(\nu_{1}\right)} \phi_{\vec{k}_{2}}^{\left(\nu_{2}\right)} \phi_{\vec{k}_{3}}^{\left(\nu_{3}\right)}
  \phi_{\vec{k}_{4}}^{\left(\nu_{4}\right)} \right\rangle_{\pm}^{\prime}
  =
  \pm i \int_{-\infty}^{\tau_{0}} \frac{\d \tau}{(-H \tau)^{\D+1}} \prod_{j=1}^{4} F_{\vec{k}_{j}, \pm}^{\left(\nu_{j}\right)}\left(\tau ; \tau_{0}\right) .
\end{equation}
Substituting \eqref{eq:MellinB2b} into this equation gives
\begin{align}
  \left\langle \phi_{\vec{k}_{1}}^{\left(\nu_{1}\right)} \phi_{\vec{k}_{2}}^{\left(\nu_{2}\right)} \phi_{\vec{k}_{3}}^{\left(\nu_{3}\right)}
  \phi_{\vec{k}_{4}}^{\left(\nu_{4}\right)} \right\rangle_{\pm}^{\prime}
  =
  \pm i H^{-\D-1} \N_4(\tau_0, k_i)
  &\int [\d s]_4 \, \rho(\bm{s},\bm{\nu})
  \prod_{j=1}^{4} e^{\delta_{\nu_j}^{\pm}(s_j)} \left( \frac{k_j}{2} \right)^{-2 s_j} \nonumber \\
  &\qquad \times
  \int_{-\infty}^{\tau_0} d\tau \, (-\tau)^{\D-1-2(s_1+s_2+s_3+s_4)} , \label{eq:Mellin4pt2}
\end{align}
where $[\d s]_N \deq \prod_{j=1}^N \int \frac{\d s_j}{2 \pi i}$ and
\begin{align}
  \rho(\bm{s},\bm{\nu}) &\deq \prod_{j=1}^{4} \Gamma\left(s_{j}+\frac{i \nu_{j}}{2}\right) \Gamma\left(s_{j}-\frac{i \nu_{j}}{2}\right) , \label{eq:rhosv} \\
  \N_4(\tau_0, k_i) &\deq \prod_{j=1}^{4} \left( \frac{k_j}{2} \right)^{i \nu_j} \N_{\nu_j}(\tau_0).
\end{align}
The momentum and time arguments which were previously trapped inside the arguments of Hankel functions are now out in the open. We can do the time integral right away,
\begin{equation}
  \int_{-\infty}^{\tau_0} \d\tau \, (-\tau)^{\D-1-2(s_1+s_2+s_3+s_4)}
  = \frac{-(-\tau_0)^{\D-2(s_1+s_2+s_3+s_4)}}{\D-2(s_1+s_2+s_3+s_4)}
  \overset{\tau_0 \to 0}{=}
  i \pi \delta \left( {\ss \frac{\D}{2}} - (s_1+s_2+s_3+s_4) \right)
  \label{eq:TimeIntegral}
\end{equation}
which converges\footnote{Usually the in-in contours are deformed slightly, such that the lower limit of the forward and return legs are $-\infty(1 \mp i\e)$, to kill the contribution from very early times. The same thing is accomplished here by constraining the real part of the Mellin variables.} for $\Re \left( \frac{\D}{2} - (s_1+s_2+s_3+s_4) \right) < 0$. Applying the delta function constraint and reorganizing a bit allows us to write \eqref{eq:Mellin4pt2} as
\begin{gather}
  \left\langle \phi_{\vec{k}_{1}}^{\left(\nu_{1}\right)} \phi_{\vec{k}_{2}}^{\left(\nu_{2}\right)} \phi_{\vec{k}_{3}}^{\left(\nu_{3}\right)}
  \phi_{\vec{k}_{4}}^{\left(\nu_{4}\right)} \right\rangle_{\pm}^{\prime}
  =
  \pm i \frac{H^{-\D-1}}{2} e^{\mp i \frac{\pi}{2}(\D+i(\nu_1+\nu_2+\nu_3+\nu_4))} \N_4(\tau_0,k_i) I(\bm{k}, \bm{\nu}) , \\
  \label{eq:MellinI}
  I(\bm{k}, \bm{\nu})
  \deq \int [\d s]_4 \, 2\pi i \, \delta \left( \frac{\D}{2} - (s_1+s_2+s_3+s_4) \right) \rho(\bm{s},\bm{\nu}) 2^D \prod_{j=1}^{4} k_j^{-2 s_j} .
\end{gather}
The full 4-point function is the sum of the contributions from the forward and return legs of the in-in contour,

  \begin{equation}
    \left\langle \phi_{\vec{k}_{1}}^{\left(\nu_{1}\right)} \phi_{\vec{k}_{2}}^{\left(\nu_{2}\right)} \phi_{\vec{k}_{3}}^{\left(\nu_{3}\right)}
    \phi_{\vec{k}_{4}}^{\left(\nu_{4}\right)} \right\rangle^{\prime}
    =
    H^{-\D-1} \sin \left(\frac{\pi}{2}(\D+i(\nu_1+\nu_2+\nu_3+\nu_4)) \right) \N_4(\tau_0,k_i) I(\bm{k}, \bm{\nu}) . \label{eq:Mellin4pt3}
  \end{equation}

\section{The anomalous dimension of \texorpdfstring{$\phi^2$}{phi squared}}
\label{sec:AnomDim}

As our first example, we compute at 1-loop the anomalous dimension of the $\phi^2$ operator, $\gamma_{\phi^2}$. This quantity was computed for the conformal mass case in \cite{Green:2020txs}. We will set up the problem for a scalar field with general mass $m$ and plug in a specific value for $m$ later.
\begin{figure}
  \centering
  \begin{tikzpicture}

      \draw[thick] (4,3) circle [radius=1];
      \filldraw[thick,draw=black,fill=white] (3,3) circle (4pt) node[left] { $\phi^2(\k,\tau_0)$ \ };
      \node[cross out, draw=black, very thick, scale=0.7] at (3,3) {};
      \filldraw[thick,draw=black,fill=white] (5,3) circle (4pt) node[right] {\ $\phi^2(\k',\tau_0)$ };
      \node[cross out, draw=black, very thick, scale=0.7] at (5,3) {};

      \node at (4,4) [above] {$\k+\p$};
      \node at (4,2) [below] {$\p$};
    \end{tikzpicture}
  \caption{\label{fig:phi2phi2corrtree} $\langle \phi^2 \phi^2 \rangle$ at tree level.}
\end{figure}
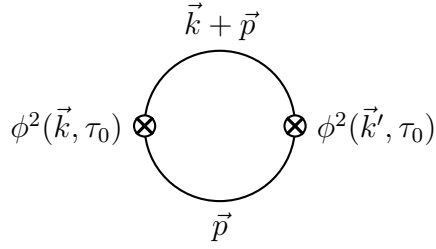
We start with the tree level expression for $\langle \phi^2 \phi^2 \rangle$ (see \figref{fig:phi2phi2corrtree}),
\begin{equation}
  \langle \phi^2 \phi^2 \rangle^\prime_{\rm tree} =
    2 \int \frac{\d^\D p}{(2 \pi)^\D} G^{(\nu)}_{\k+\p}(\tau_0) G^{(\nu)}_{\p}(\tau_0)
    \label{eq:phi2phi2treegeneral1}
\end{equation}
where $G^{(\nu)}$ is the late time two point function
\begin{equation}
  G^{(\nu)}_{\k}(\tau_0)
    = \lim_{\tau_0 \to 0} \left\langle \phi_{\k}^{(\nu)} (\tau_0) \phi_{\k'}^{(\nu)} (\tau_0) \right\rangle^{\prime}
    = \frac{H^{\D-1}}{4 \pi} \Gamma(-i \nu)^2 (-\tau_0)^{\D + 2i\nu} \left (\frac{k}{2}\right)^{2i\nu} .
  \label{eq:LateTime2ptFn}
\end{equation}

\noindent We could, for instance, arrive at this expression by taking the limit $\tau \to \tau_0 \sim 0$ in \eqref{eq:MellinB2b} in which case the integral is dominated by the pole at\footnote{The integrand in \eqref{eq:MellinB2b} also has poles $s_\star = \frac{i \nu}{2} + n$ but these produce analytic terms in $k$ which don't give rise to long-distance correlations. See \cite{Sleight:2019mgd}.} $s_\star= -\frac{i \nu}{2}$.
%
%
%
The momentum integral in \eqref{eq:phi2phi2treegeneral1} can be computed using the convolution theorem (cf.\ \eqref{eq:ImomConvolution}). The result is
\begin{equation}
  \langle \phi^2 \phi^2 \rangle^\prime_{\rm tree}
    = \frac{2}{(4 \pi)^{\frac{\D}{2}+2}}
    \frac{
      \Gamma(-{\ss \frac{\D}{2}} - 2i\nu) \Gamma({\ss \frac{\D}{2}} + i\nu)^2 \Gamma(-i\nu)^2
      }{
      \Gamma(\D + 2i\nu)
      } \,
      H^{2\D-2} (-\tau_0)^{2\D + 4i\nu} k^{\D + 4i\nu} .
      \label{eq:phi2phi2treegeneral2}
\end{equation}
\begin{figure}[ht]
  \begin{tikzpicture}

      \begin{scope}[xshift=-1.5cm]
         \draw[thick] (-0.5,4) -- (4.5,4) node[right,yshift=-3]{$\tau_0$};
         \draw[thick] (0,4) -- (2,2) -- (4,4);
         \draw[thick] (1.5,4) -- (2,2) -- (2.5,4);
         \filldraw (2,2) circle (2pt) node[below] {$\tau$};

         \filldraw (0,4) circle (1pt) node[above] {$\k+\p_1$};
         \filldraw (1.5,4) circle (1pt) node[above] {$\p_1$};
         \filldraw (2.5,4) circle (1pt) node[above] {$\p_2$};
         \filldraw (4,4) circle (1pt) node[above] {$\k+\p_2$};
         \draw[->,thick] ({1.1 + 0.3*cos(120)},{3.5 + 0.3*sin(120)}) arc[radius = 3mm, start angle=120, end angle=420];
         \node at (1.1,3.5) {$\p_1$};
         \draw[->,thick] ({2.9 + 0.3*cos(120)},{3.5 + 0.3*sin(120)}) arc[radius = 3mm, start angle=120, end angle=420];
         \node at (2.9,3.5) {$\p_2$};
      \end{scope}

      \draw[thick] (7,3) circle [radius=1];
      \draw[thick] (9,3) circle [radius=1];
      \filldraw (8,3) circle (2pt) node[right] {$\tau$};
      \filldraw[thick,draw=black,fill=white] (6,3) circle (4pt) node[left] { $\phi^2(\k,\tau_0)$ \ };
      \node[cross out, draw=black, very thick, scale=0.7] at (6,3) {};
      \filldraw[thick,draw=black,fill=white] (10,3) circle (4pt) node[right] {\ $\phi^2(\k',\tau_0)$ };
      \node[cross out, draw=black, very thick, scale=0.7] at (10,3) {};

      \node at (7,4) [above] {$\k+\p_1$};
      \node at (9,4) [above] {$\k+\p_2$};
      \node at (7,2) [below] {$\p_1$};
      \node at (9,2) [below] {$\p_2$};

        \draw[->, very thick] (2.5,3) -- (3.7,3);

      \draw[->,thick] (-2,2.25) -- (-2,3.5);
      \node[below] at (-2,2.25) {time};
    \end{tikzpicture}
  \caption{\label{fig:phi2phi2firstordercorr}The first order correction to $\langle \phi^2 \phi^2 \rangle$.}
\end{figure}
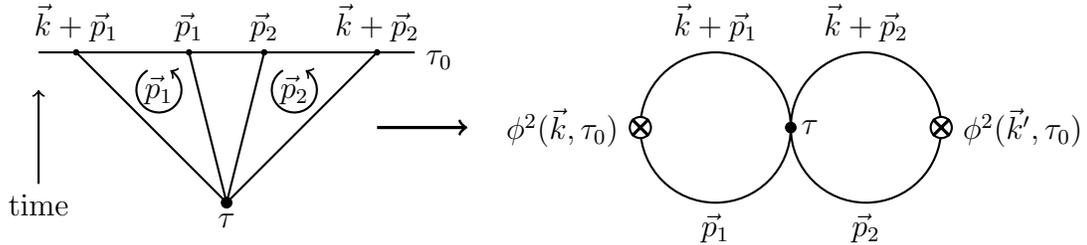
Next, we calculate the first order correction $\langle \phi^2 \phi^2 \rangle_\lambda$ which is obtained by contracting the legs of the 4-point correlation function \eqref{eq:Mellin4pt3} as shown in \figref{fig:phi2phi2firstordercorr}.
This corresponds to setting $\k_1 = \k+\p_1, \k_2 = \p_1, \k_3 = \k+\p_2$ and $\k_4 = \p_2$ in \eqref{eq:Mellin4pt3} and integrating over $\p_1$ and $\p_2$,
\begin{align}
  \langle \phi^2 \phi^2 \rangle^\prime_{\lambda}
  &= \int \frac{\d^\D p_1}{(2 \pi)^\D} \frac{\d^\D p_2}{(2 \pi)^\D}
  \left\langle
    \phi_{\k+\p_1}^{\left(\nu_{1}\right)}
    \phi_{\p_1}^{\left(\nu_{2}\right)}
    \phi_{\k+\p_2}^{\left(\nu_{3}\right)}
    \phi_{\p_2}^{\left(\nu_{4}\right)}
  \right\rangle^{\prime} \nonumber \\[0.75em]
  &=
  C \int [\d s]_4 \, 2\pi i \, \delta \left( {\ss \frac{\D}{2}} - (s_1+s_2+s_3+s_4) \right) \rho(\bm{s},\bm{\nu}) \,2^\D I_{\rm spec} (k, \bm{s},\bm{\nu}) .
  \label{eq:phi2phi2firstorder1}
\end{align}
Here, $I_{\rm spec}$ is the momentum integral and we have collected the prefactors into the constant $C$,
\begin{align}
I_{\rm spec}(k, \bm{s},\bm{\nu}) &=
  \int \frac{\d^D p_1}{(2 \pi)^D} \frac{\d^D p_2}{(2 \pi)^D}
  \frac{1}{|\k + \p_1|^{2s_1-i\nu_1}} \frac{1}{p_1^{2s_2-i\nu_2}}
  \frac{1}{|\k + \p_2|^{2s_3-i\nu_3}} \frac{1}{p_2^{2s_4-i\nu_4}} \label{eq:Ispec} \\
  C &= H^{-\D-1} \sin \left( \frac{\pi}{2}(D+i(\nu_1+\nu_2+\nu_3+\nu_4)) \right) \left( \prod_{j=1}^{4} \frac{1}{2^{i \nu_j}} \N_{\nu_j}(\tau_0) \right) .
  \label{eq:4ptprefactor}
\end{align}
We keep the masses distinct for now, leaving open the possibility of using the $\nu_j$'s to regulate the divergences in \eqref{eq:phi2phi2firstorder1}. The momentum integral \eqref{eq:Ispec} clearly factorizes into two integrals of the form
\begin{equation}
  I_{\rm mom} (a,b) = \int \frac{\d^\D p}{(2 \pi)^D}
  \frac{1}{|\k + \p \,|^{a}} \frac{1}{p^{b}} .
\end{equation}
This presents us with the opportunity to introduce an analytic continuation parameter. To see how, we first note that $I_{\rm mom}$ is just the convolution of the function $\tilde f_n(\p) = 1/p^n$ with itself,
\begin{equation}
I_{\rm mom} (a,b)
  = \int \frac{\d^D p}{(2 \pi)^D} \tilde f_a(\k + \p) \tilde f_b(\p)
  = \int \d^\D y e^{-i \k \cdot \y} f_a(\y) f_b(\y) ,
  \label{eq:ImomConvolution}
\end{equation}
where $f_n(\y)$ is given by the radial Fourier transform
\begin{equation}
  f_n(\y)
    = \int \frac{\d^D p}{(2 \pi)^D} \frac{e^{i \p \cdot \y}}{p^n}
    = \frac{1}{2^n \pi^{D/2}} \frac{\Gamma \left( \frac{D-n}{2} \right)}{\Gamma \left( \frac{n}{2} \right)} y^{n-D} . \label{eq:radialfourierxfrm}
\end{equation}
Before substituting this into \eqref{eq:ImomConvolution} we change the integral measure $\d^\D y \to \d^{\bar \D} y$ with the promise that we'll take $\bar \D \to \D$ at the end. We will also introduce a factor of $(-\tau_0)^{\D-\bar \D}$ in front to preserve the overall dimensions and to ensure that the momenta $\k$ appears as the product $-k \tau_0$ in the final answer.
This way it remains invariant under the rescaling $k \to \rho^{-1} k$ and $a(\tau) \to \rho^{-1} a(\tau)$ \cite{Green:2020txs}. With these changes
\begin{align}
  \bar{I}_{\rm mom} (a,b)
    &= \frac{1}{2^{a+b} \pi^{\D}} \frac{\Gamma \left( \frac{\D-a}{2} \right) \Gamma \left( \frac{\D-b}{2} \right)}{\Gamma \left( \frac{a}{2} \right) \Gamma \left( \frac{b}{2} \right)} \
    (-\tau_0)^{\D-\bar{\D}} \int \d^{\bar{\D}} y \frac{e^{-i \k \cdot \y}}{y^{2\D-a-b}} \nonumber \\[0.75em]
    &\overset{\eqref{eq:radialfourierxfrm}}{=}
    \frac{1}{(4 \pi)^{\frac{2\D-\bar{\D}}{2}}}
      \frac{
        \Gamma \left( \frac{\D-a}{2} \right) \Gamma \left( \frac{\D-b}{2} \right)
      }{
        \Gamma \left( \frac{a}{2} \right) \Gamma \left( \frac{b}{2} \right)
      }
      \frac{
        \Gamma \left( \frac{a+b}{2} - \frac{2\D-\bar{\D}}{2} \right)
      }{
        \Gamma \left( \D-\frac{a+b}{2} \right)
      }
      k^{\D-a-b} (-k \tau_0)^{\D-\bar{\D}} \label{eq:Imom}
\end{align}
where we have introduced a bar above $\bar{I}_{\rm mom}$ to indicate that we have tampered with the dimension $\D$ at an intermediate step. The difference $\D - \bar \D$ can be used as an analytic continuation parameter. We introduce this parameter into one of the momentum integrals in \eqref{eq:Ispec} by writing
\begin{align}
  I_{\rm spec}
    &= \bar{I}_{\rm mom} (2s_1-i\nu_1, 2s_2-i\nu_2) \times I_{\rm mom} (2s_3-i\nu_3, 2s_4-i\nu_4)
\end{align}
with the other momentum integral obtained by setting $\D=\bar D$ in \eqref{eq:Imom}. We may now substitute this into \eqref{eq:phi2phi2firstorder1} and integrate over $s_4$ to apply the delta function. The result is
%
\begin{align}
  \langle &\phi^2 \phi^2 \rangle^\prime_{\lambda}
  =
  C \frac{2^\D}{(4 \pi)^{ \frac{3\D-\bar{\D}}{2} }} (-k \tau_0)^{\D - \bar{\D}} \, k^{D + i\sum_j \nu_j} \nonumber \\
  &\times  \int [\d s]_3 \,
  \Gamma \left( {\ss \frac{\D}{2}} - s_1 - s_2 - s_3 + {\ss \frac{i\nu_{4}}{2}} \right)
  \Gamma \left( s_1 + s_2 + s_3  + {\ss \frac{i\nu_4}{2}} \right)
  \prod_{j=1}^{3}
    \Gamma \left( s_{j} + {\ss \frac{i \nu_{j}}{2}} \right)
    \Gamma \left( {\ss \frac{\D}{2}} - s_j + {\ss \frac{i\nu_j}{2}} \right) \nonumber \\
  &\qquad \times
  \frac{
    \Gamma \left( \frac{\bar{D}}{2} - D + s_1 + s_2 - \frac{i \nu_1}{2} - \frac{i \nu_2}{2} \right)
  }{
    \Gamma \left( D - s_1 - s_2 + \frac{i \nu_1}{2} + \frac{i \nu_2}{2} \right)
  }
  \frac{
    \Gamma \left( -s_1 - s_2 - \frac{i \nu_3}{2} - \frac{i \nu_4}{2} \right)
  }{
    \Gamma \left( \frac{D}{2} + s_1 + s_2 + \frac{i \nu_3}{2} + \frac{i \nu_4}{2} \right)
  }. \label{eq:phi2phi2firstorder2}
\end{align}
%
The divergences are completely encoded in the 3-fold MB integral. To proceed we will need to plug in values for $\D$ and $\nu_j$.

\subsection{The conformal mass case}
\label{sec:AnomDimConformal}

We will first compute $\gamma_{\phi^2}$ for a scalar field with conformal mass $m^2 = 2H^2$ in $\D=3$ dimensions. This corresponds to $\nu = \frac{i}{2}$. Substituting these numbers into \eqref{eq:phi2phi2treegeneral2} we obtain the tree level contribution
\begin{equation}
  \langle \phi^2 \phi^2 \rangle^\prime_{\rm tree} = -\frac{1}{8 \pi^2} (-H \tau_0)^4 k .
  \label{eq:phi2phi2treeconformal}
\end{equation}
The first order correction follows from setting
\begin{equation}
  \begin{aligned}
    \nu_2 &\to \frac{i}{2} - i 2 \alpha \\
    \nu_j &\to \frac{i}{2}, \quad j \neq 2 \\
    \bar{\D} &\to 3 + 2 \delta
  \end{aligned} \label{eq:ConformalNusAndDbar}
\end{equation}
in \eqref{eq:phi2phi2firstorder2}. We have introduced the parameters $\alpha$ and $\delta$ to make the MB integral well-defined, that is, to remove any overlaps of left/right poles as required by the Mellin contour prescription\footnote{It is possible to remove all pole overlaps in this case with just one parameter, as discussed in \secref{sec:AnomAlternateParam}. We chose to use two here to illustrate some aspects of using multiple parameters.}.
These are the parameters denoted by $\e_k$ in \eqref{eq:GammaFnArgumentsModified}. The actual integral, with $\alpha=0=\delta$ , will be defined by analytically continuing in these parameters. With these substitutions
\begin{equation}
  \langle \phi^2 \phi^2 \rangle^\prime_{\lambda} =
    C \frac{8}{(4 \pi)^{3-\delta}}
    k^{1+2\alpha} (-k \tau_0)^{-2\delta}
    \int [\d s]_3 \, \frac{\Gamma_1 \Gamma_2  \Gamma_3 \Gamma_4 \Gamma_5 \Gamma_6 \Gamma_7 \Gamma_8 \Gamma_9 \Gamma_{10}}{\Gamma_{11} \Gamma_{12}} ,
    \label{eq:phi2phi2firstorder3}
\end{equation}
where we have labelled the $\Gamma$-functions,
\begin{equation}
  \begin{aligned}
    \Gamma_1 &= \Gamma(-{\ss \frac{1}{4}} + s_1 ) \\
    \Gamma_2 &= \Gamma( {\ss \frac{5}{4}} - s_1 ) \\
    \Gamma_3 &= \Gamma(-{\ss \frac{1}{4}} + s_2 + \alpha ) \\
    \Gamma_4 &= \Gamma( {\ss \frac{5}{4}} - s_2 + \alpha ) \\
    \Gamma_5 &= \Gamma(-{\ss \frac{1}{4}} + s_3 ) \\
    \Gamma_6 &= \Gamma( {\ss \frac{5}{4}} - s_3 ) \\
    \Gamma_7 &= \Gamma(-{\ss \frac{1}{4}} + s_1 + s_2 + s_3 ) \\
    \Gamma_8 &= \Gamma( {\ss \frac{5}{4}} - s_1 - s_2 - s_3 ) \\
    \Gamma_9 &= \Gamma( {\ss \frac{1}{2}} - s_1 - s_2 ) \\
    \Gamma_{10} &= \Gamma(-1 + s_1 + s_2 - \alpha + \delta ) \\
    \Gamma_{11} &= \Gamma(1 + s_1 + s_2 ) \\
    \Gamma_{12} &= \Gamma( {\ss \frac{5}{2}} - s_1 - s_2 + \alpha ) .
  \end{aligned}
  \label{eq:AnomGammas}
\end{equation}
The evaluation of this integral is discussed in detail below. However, most parts of the calculation can be done using the \texttt{MB} Mathematica package by Czakon \cite{Czakon:2005rk}.

\subsubsection{Choosing the contours}
\label{sec:ChoosingContoursAnom}

We shall focus on the MB integral
\begin{equation}
  K = \int [\d s]_3 \, \frac{\Gamma_1 \Gamma_2  \Gamma_3 \Gamma_4 \Gamma_5 \Gamma_6 \Gamma_7 \Gamma_8 \Gamma_9 \Gamma_{10}}{\Gamma_{11} \Gamma_{12}}.
  \label{eq:AnomK}
\end{equation}
We will integrate this over straight line contours. The integral is well-defined if we can choose contours such that \eqref{eq:MellinContourPrescription} is satisfied. This gives the following inequalities:
\begin{equation}
  \begin{aligned}
    U_{1,2}(\C) > 0 &\implies \frac{1}{4} < s_1^\C < \frac{5}{4} \\
    U_{3,4}(\C) > 0 &\implies \frac{1}{4} - \alpha < s_2^\C < \frac{5}{4} + \alpha \\
    U_{5,6}(\C) > 0 &\implies \frac{1}{4} < s_3^\C < \frac{5}{4} \\
    U_{7,8}(\C) > 0 &\implies \frac{1}{4} < s_1^\C + s_2^\C + s_3^\C < \frac{5}{4} \\
    U_{9}(\C) > 0 &\implies s_1^\C + s_2^\C < \frac{1}{2} \\
    U_{10}(\C) > 0 &\implies s_1^\C + s_2^\C > 1 + \alpha - \delta .
  \end{aligned}
  \label{eq:AnomConstraints}
\end{equation}
These inequalities can be combined to give the condition
\begin{equation}
  {\rm max}\left( \frac{1}{4} , \, \frac{3}{4} - \alpha, \frac{5}{4} + \alpha - \delta \right)
  < s_1^\C + s_2^\C + s_3^\C <
  {\rm min}\left( \frac{5}{4}, \, \frac{7}{4}, \, \frac{15}{4} + \alpha \right) .
  \label{eq:AnomConstraintCombined1}
\end{equation}
The contours $s_1^\C$ and $s_2^\C$ have another constraint which follows from $U_{1,3,9}(\C) > 0$,
\begin{equation}
 \frac{1}{2} - \alpha < s_1^\C + s_2^\C < \frac{1}{2} .
 \label{eq:AnomConstraintCombined2}
\end{equation}
This condition can only be satisfied if $\alpha > 0$. This means that for sufficiently small $\alpha$ and $\delta$, \eqref{eq:AnomConstraintCombined1} becomes
\begin{equation}
  \frac{5}{4} + \alpha - \delta < s_1^\C + s_2^\C + s_3^\C < \frac{5}{4} ,
  \label{eq:AnomConstraintCombined3}
\end{equation}
%
which requires $\delta > \alpha$.
The inequalities \eqref{eq:AnomConstraintCombined2} and \eqref{eq:AnomConstraintCombined3} indicate that the contours are pinched, with $\alpha$ and $\delta-\alpha$ being the width of the pinches (cf. \figref{fig:MellinContour}). We will see shortly how these pinches manifest as the divergences of \eqref{eq:AnomK}.

%

The conditions \eqref{eq:AnomConstraints} can be satisfied by choosing $\alpha=0.1, \delta=0.7, s_1^\C = s_3^\C = 0.3$ and $s_2^\C = 0.17$. This choice is by no means unique.


\subsubsection{Analytic continuation}
\label{sec:AnomAC}

With the contours fixed we start at $\alpha=0.1, \delta=0.7$ and analytically continue to the region around $\alpha, \delta \sim 0$. As we decrease $\alpha$ and $\delta$ we keep track of the poles that cross the straight line contours and add the residues at those poles to the integral in \eqref{eq:AnomK}, now evaluated at a smaller value of the parameters $\alpha$ and $\delta$. These residue terms may have poles in the remaining variables which can cross other contours as the parameters are decreased further. Iterating this process we end up with a collection of residue terms, plus the original four fold integral. Some of the residue terms contain factors that diverge as $\alpha,\delta \to 0$. All remaining integrals can simply be expanded in $\alpha,\delta$ under the integral sign since there are no further pole crossings.

The present example is simple enough that we end up with only four residue terms. Decreasing to $\alpha=0.08$ and $\delta=0.56$, the first poles to cross the contours are at $s_2 = \frac{1}{4}-\alpha$ and $s_2 = 1 - s_1 + \alpha -\delta$ which are the zeroth poles of $\Gamma_3$ and $\Gamma_{10}$.
Following the pattern in \eqref{eq:SeparatingADivergence1} and \eqref{eq:RudyBreakup}, we can write
\begin{equation}
  K \to K_3 + K_{10} + K,
\end{equation}
where the $K$ on the r.h.s.\ is evaluated at the new values of $\alpha$ and $\delta$, and
\begin{align}
  K_3 &= \int \frac{\d s_1}{2 \pi i} \frac{\d s_3}{2 \pi i}
    \frac{\Gamma_1 \Gamma_2  \ \ \Gamma_4 \Gamma_5 \Gamma_6 \Gamma_7 \Gamma_8 \Gamma_9 \Gamma_{10}}{\Gamma_{11} \Gamma_{12}}
    \bigg \rvert_{
      \substack{s_2 = \frac{1}{4} - \alpha}
    } , \\[0.75em]
  K_{10} &= \int \frac{\d s_1}{2 \pi i} \frac{\d s_3}{2 \pi i}
    \frac{\Gamma_1 \Gamma_2  \Gamma_3 \Gamma_4 \Gamma_5 \Gamma_6 \Gamma_7 \Gamma_8 \Gamma_9 \ \ }{\Gamma_{11} \Gamma_{12}}
    \bigg \rvert_{
      \substack{s_2 = 1 - s_1 + \alpha -\delta}
    } .
\end{align}
The $\Gamma$ functions of $K_3$ and $K_{10}$ will have a different dependence on $\alpha$ and $\delta$ once $s_2$ is eliminated. Reducing the parameters further we find that the zeroth pole of $\Gamma_9$, at $s_1 = \frac{1}{4}+\alpha$, crosses over in $K_3$ prompting us to write $K_3 \to K_{39} + K_3$.
At the same time the pole $s_3 = \frac{1}{4}-\alpha+\delta$ of $\Gamma_8$ crosses over in $K_{10}$ leading to the split $K_{10} \to K_{10,8} + K_{10}$. There are no further pole crossings as $\alpha,\delta \to 0$. All in all, we end up with
\begin{equation}
  K \to K_{39} + K_{10,8} + K_3 + K_{10} + K . \label{eq:AnomKBreakup}
\end{equation}
The term $K_{39}$ is
\begin{equation}
  \frac{
    2 \Gamma (\alpha ) \Gamma (1-\alpha )
    \Gamma (2 \alpha + 1) \Gamma(-\frac{1}{2} - \alpha + \delta)
   }{
   \sqrt{\pi} \Gamma (\alpha +2)
   }
   \int_{0.3 - i \infty}^{0.3 + i \infty} \frac{\d s_3}{2 \pi i}
   \Gamma( {\ss \frac{3}{4}} - s_3)
   \Gamma( {\ss \frac{5}{4}} - s_3)
   \Gamma(-{\ss \frac{1}{4}} + s_3)
   \Gamma( {\ss \frac{1}{4}} + s_3) .
   \label{eq:K39Integral}
\end{equation}
The left/right poles in the $s_3$ integral are separated by the contour. Therefore we can evaluate this integral using Barnes's first lemma\footnote{Barnes's first lemma states
\begin{equation}
\begin{aligned}
  \int_{-i \infty}^{+i \infty} \frac{\d z}{2 \pi i}
    \Gamma &(\lambda_{1}+z) \Gamma(\lambda_{2}+z) \Gamma(\lambda_{3}-z) \Gamma(\lambda_{4}-z)
  = \frac{
    \Gamma\left(\lambda_{1}+\lambda_{3}\right) \Gamma\left(\lambda_{1}+\lambda_{4}\right) \Gamma\left(\lambda_{2}+\lambda_{3}\right) \Gamma\left(\lambda_{2}+\lambda_{4}\right)
  }{
    \Gamma\left(\lambda_{1}+\lambda_{2}+\lambda_{3}+\lambda_{4}\right)
  } , \label{eq:B1L}
\end{aligned}
\end{equation}
where the contour is a vertical line that separates the left/right poles of the $\Gamma$ functions in the integrand.
} to obtain
\begin{equation}
  K_{39} = - 2 \pi \Gamma(\alpha) + O(1) .
\end{equation}
Next, the integral $K_{10,8}$ is
\begin{align}
  K_{10,8}
    &=
    \frac{
      \Gamma ( \delta - \alpha )
      \Gamma ( 1 + \alpha - \delta )
      \Gamma (-{\ss \frac{1}{2}} - \alpha + \delta )
    }{
      \Gamma (\frac{3}{2} + \delta )
      \Gamma ( 2 + \alpha - \delta )
    } \nonumber \\[0.75em]
    &\qquad \times
    \int_{0.3 - i \infty}^{0.3 + i \infty} \frac{\d s_1}{2 \pi i}
    \Gamma ( {\ss \frac{5}{4}} - s_1)
    \Gamma (-{\ss \frac{1}{4}} + s_1)
    \Gamma ({\ss \frac{1}{4}} + s_1 + \delta )
    \Gamma ({\ss \frac{3}{4}} - s_1 + 2\alpha - \delta ) .
    \label{eq:K108Integral}
\end{align}
%
The poles of the of the $s_1$ integral are well separated by the straight line contour around $\alpha, \delta \sim 0$. Therefore we can once again use \eqref{eq:B1L} to obtain
\begin{equation}
  K_{10,8} = - 2 \pi \Gamma( \delta - \alpha ) + O(1) .
\end{equation}
The remaining terms $K_3$, $K_{10}$ and $K$ in \eqref{eq:AnomKBreakup} are at most\footnote{$O(1)$ terms are those which do not blow up as the analytic continutation parameters approach 0.} $O(1)$ and can be ignored. Putting the pieces together we have
\begin{equation}
  K = -2 \pi \left[ \Gamma(\alpha) + \Gamma( \delta - \alpha ) \right] + O(1) ,
  \label{eq:AnomKResult}
\end{equation}
which completes the evaluation of \eqref{eq:AnomK}. Substituting the above result into \eqref{eq:phi2phi2firstorder3},
\begin{align}
  \langle \phi^2 \phi^2 \rangle^\prime_{\lambda} &=
    \lim_{\alpha, \delta \to 0}
    C \frac{8}{(4 \pi)^{3-\delta}}
    k^{1+2\alpha} (-k \tau_0)^{-2\delta} \times
    -2 \pi \left[ \Gamma(\alpha) + \Gamma( \delta - \alpha ) \right] .
\end{align}
We evaluate the prefactor $C$ defined in \eqref{eq:4ptprefactor} with the $\nu_j$ from \eqref{eq:ConformalNusAndDbar}
and\footnote{The number of space dimensions in \eqref{eq:4ptprefactor} is $D=3$ and not the $\bar{\D}=3+2\delta$ introduced in \eqref{eq:Imom}.} $\D=3$,
\begin{equation}
  C = \frac{H^4}{2^{6+2\alpha} \pi^{5/2}}
    \cos ( \pi \alpha )
    \Gamma ({\ss \frac{1}{2}} - 2\alpha )
    (-\tau_0)^{4+2\alpha}
    \sim \frac{H^4}{64 \pi^2} (-\tau_0)^{4+2\alpha} .
\end{equation}
Then, at leading order in $\alpha$ and $\delta$ we have
\begin{equation}
  \langle \phi^2 \phi^2 \rangle^\prime_{\lambda}
    = \frac{(-H \tau_0)^4}{512 \pi^5} k \times
      -2 \pi \left[ \frac{1}{\alpha} + \frac{1}{\delta-\alpha} - 2 \gamma_E + \cdots \right] (-k \tau_0)^{2\alpha-2\delta} .
  \label{eq:AnomDimTwoParamResult}
\end{equation}
As expected the divergences are directly related to the width of the pinches. However, we do not know the exact relationship between $\alpha$ and $\delta$ except that they are both infinitesimal and $\delta > \alpha$ (this condition, which follows from \eqref{eq:AnomConstraintCombined3}, fixes the sign of the $\frac{1}{\delta-\alpha}$ divergence and we must adhere to it throughout the analytic continuation process). It is `reasonable' to make the width of the pinches equal by setting $\delta - \alpha = \alpha$. Then,
\vskip 5pt
\begin{eBox}
  \begin{equation}
    \langle \phi^2 \phi^2 \rangle^\prime_{\lambda}
      = -\frac{(-H \tau_0)^4}{128 \pi^4} k
        \left[ \frac{1}{\alpha} - 2 \log (-k \tau_0) + \cdots \right] .
    \label{eq:Anom1stOrderResult}
  \end{equation}
\end{eBox}
%

\subsubsection{An alternate parameterization}
\label{sec:AnomAlternateParam}

It is possible to define \eqref{eq:AnomK} with a single parameter instead of the $\alpha$ and $\delta$ we introduced in \eqref{eq:ConformalNusAndDbar}. This time we will set all $\nu_j \to \frac{i}{2}$ and $\bar{\D}=\D=3$ but modify the time integral \eqref{eq:TimeIntegral} to
\begin{equation}
  (-\tau_0)^{-2\e} \int_{-\infty}^{\tau_0} \d\tau \, (-\tau)^{\D-1-2(s_1+s_2+s_3+s_4)+2\e}
  \overset{\tau_0 \to 0}{=}
  (-\tau_0)^{-2\e} \times i \pi \delta \left( {\ss \frac{\D}{2}} - (s_1+s_2+s_3+s_4) + \e \right)
  \label{eq:TimeIntegralWithParameter}
\end{equation}
where $\e$ is the new parameter in which we will analytically continue. We have used a factor of $(-\tau_0)^{-2\e}$ to keep the dimensions correct, just as we did in \eqref{eq:Imom}. Introducing these changes into \eqref{eq:phi2phi2firstorder2} and repeating the calculation we get
\begin{equation}
  \langle \phi^2 \phi^2 \rangle^\prime_{\lambda}
    = -\frac{(-H \tau_0)^4}{128 \pi^4} k
      \left[ \frac{1}{\e} - 2 \log (-k \tau_0) + \cdots \right] ,
\end{equation}
in agreement with \eqref{eq:Anom1stOrderResult}. Physically, this regulator deforms the background by a small amount, and one might worry that it breaks the underlying dS symmetries. However, the degree of this breaking is controlled by the small number $\e$, as evidenced by the slight non-conservation of $s$ in \eqref{eq:Anom1stOrderResult}. If we remove the $1/\e$ piece with a counterterm, we are left with an answer that is independent of $\e$ and is valid in the fully symmetric $\e \to 0$ limit.

Furthermore, there is no ambiguity about the relationship between analytic continuation parameters here since there is only one of them. We also see that it was correct to require the width of the pinches match, for, any other choice would have led to a different answer. This piece of insight will be useful in the next example where it is impossible to make the MB integral well-defined with just one parameter.

\subsubsection{Dynamical renormalization}

The expression \eqref{eq:Anom1stOrderResult}, with its $1/\alpha$ pole and the associated log, resembles a standard 1-loop result in flat space, computed by continuing the number of spacetime dimensions\footnote{We can push the analogy further, to see that resumming the logs at all orders will simply change the scaling dimension of the $\phi^2$ operator (cf. \eqref{eq:phi2phi2Resummed}). The parallel between DRG and RG is explored in greater detail in \cite{Green:2020txs}.}.
The term $\log(-k\tau_0) \equiv \log(k/(aH))$ blows up at late times, $\tau_0 \to 0$, jeopardizing the validity of the perturbation series. This indicates that our series expansion was too simple-minded to begin with, and more careful treatment is required to handle these secular growth terms. The Dynamical Renormalization Group (DRG) \cite{Burgess_2010,Chen:1995ena} provides the required fix.
To review quickly, we start by combining \eqref{eq:phi2phi2treeconformal} and \eqref{eq:Anom1stOrderResult},
\begin{align}
  \langle \phi^2 \phi^2 \rangle^\prime
  &= \langle \phi^2 \phi^2 \rangle^\prime_{\rm tree} - \lambda \langle \phi^2 \phi^2 \rangle^\prime_{\lambda} + O(\lambda^2) \nonumber \\[0.75em]
  &= -\frac{1}{8 \pi^2} (-H \tau_0)^4 k
    \left[
      1 - \frac{\lambda}{16\pi^2} \left( \frac{1}{\alpha} - 2 \log(- k \tau_0) + \cdots \right) + O(\lambda^2)
    \right] .
  \label{eq:phi2phi2BeforeResum}
\end{align}
We now remove the divergence with a counterterm, which also introduces a new ratio of scales $k_\star/(aH)_\star$ (see \cite{Burgess_2010,Green:2020txs} for more details). Noticing that the correlation function must be independent of this ratio, we obtain a differential equation for $\langle \phi^2 \phi^2 \rangle$, the solution of which resums the secular logs to
\begin{align}
    \langle \phi^2 \phi^2 \rangle^\prime
    &=
    -\frac{1}{8 \pi^2} (-H \tau_0)^4 \, k
    \exp \left(
      \frac{\lambda}{8 \pi^2}
      \log ( -k \tau_0 ) + \dots
    \right) (1+ \dots) \nonumber \\
    &= -\frac{1}{8 \pi^2} (-H \tau_0)^4 \, k \, ( -k \tau_0 )^{2 \gamma_{\phi^2}} \, (1 + O(\lambda^2)) \nonumber \\
    &= -\frac{1}{8 \pi^2} \frac{H^{-2\gamma_{\phi^2}}}{a(\tau_0)^{2\Delta_{\phi^2} + 2 \gamma_{\phi^2}}} \, k^{1+2\gamma_{\phi^2}} \, (1 + O(\lambda^2)) .
    \label{eq:phi2phi2Resummed}
\end{align}
Thus, treating the secular dependence with DRG induces an anomalous dimension for the $\phi^2$ operator, namely
\begin{equation}
  \gamma_{\phi^2} = \frac{\lambda}{16 \pi^2} .
\end{equation}
%
%
This anomalous scaling is a product of subhorizon effects, as we demonstrate in \secref{sec:DivergenceStructure}. For now, we note that the treatment of secular growth in dS closely resembles the handling of UV divergences in flat space.
This connection holds for all scalar fields of general mass, except when the field is massless; massless scalars in dS have no flat space analog.

\subsection{\texorpdfstring{$\gamma_{\phi^2}$}{Anomalous dimension} for other masses}
\label{sec:AnomDimOtherMasses}

{
\newfloatcommand{capbtabbox}{table}[][\FBwidth]
\begin{figure}
\begin{floatrow}
\CenterFloatBoxes
  \ffigbox[\FBwidth]
  {\includegraphics[scale=0.63]{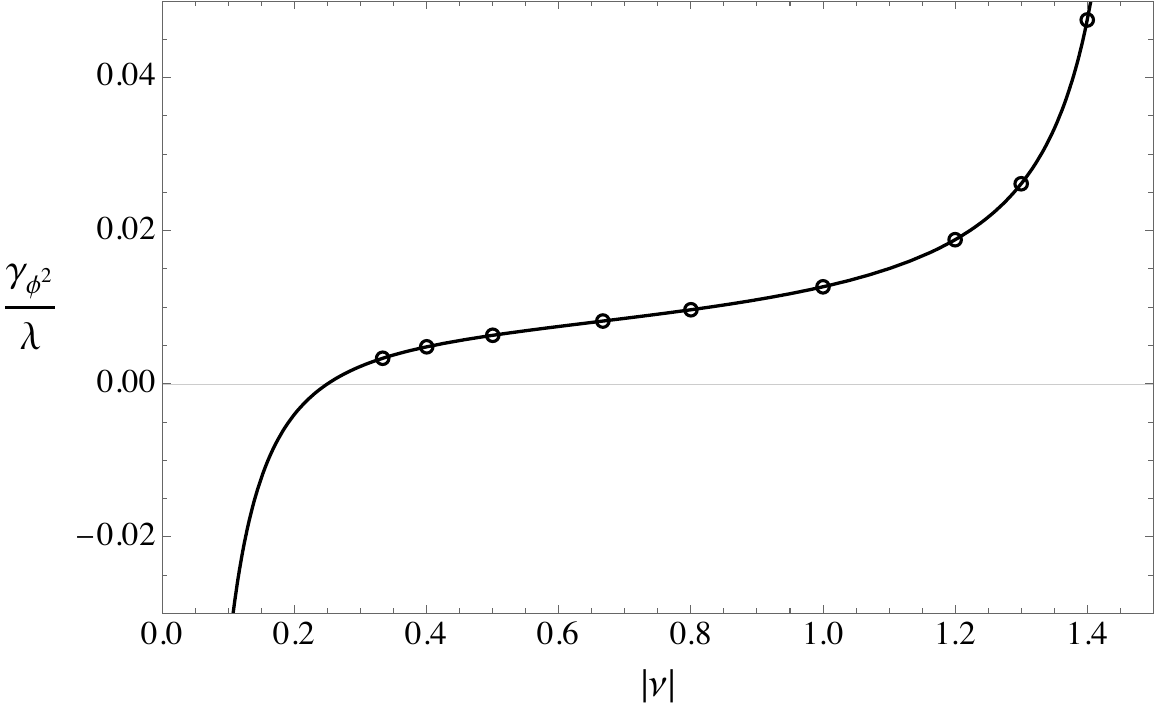}}
  {\caption{\label{fig:AnomDimensionsVsMasses}Anomalous dimension of $\phi^2$ for different masses. Larger values of $|\nu|$ correspond to smaller masses (see \eqref{eq:nuDefn}). The solid black line is from the closed form expression \eqref{eq:AnomDimClosedForm}.
  }}
  \killfloatstyle
  \capbtabbox
  {
  \bgroup
    \def\arraystretch{1.5}%
      \begin{tabular}{|c|c|}
      \hline
       $|\nu|$ & $\gamma_{\phi^2}/\lambda$ \\ \hline
       $1/3$ & $0.00333$ \\ \hline
       $2/5$ & $0.00483$ \\ \hline
       $1/2$ & $\frac{1}{16 \pi^2}$  \\ \hline
       $2/3$ & $0.00819$ \\ \hline
       $4/5$ & $0.00967$ \\ \hline
       $1$ & $\frac{1}{8 \pi^2}$ \\ \hline
       $6/5$ & $0.01882$ \\ \hline
       $13/10$ & $0.02612$ \\ \hline
       $7/5$ & $0.04753$ \\ \hline
      \end{tabular}
    \egroup
  }{
  \caption{\label{tbl:AnomDimensionsVsMasses}}
  }
\end{floatrow}
\end{figure}
}
The calculation above was repeated for a few other light scalars with masses in the range $0 < m^2 < \left(\frac{\D}{2}\right)^2 H^2$. The results are summarized in Table \ref{tbl:AnomDimensionsVsMasses}, and plotted in \figref{fig:AnomDimensionsVsMasses}.
An interesting feature of this graph is the sharp rise in the value of $\gamma_{\phi^2}$ as we approach the massless limit, $\nu \to i 3/2$. This indicates another divergence creeping in as we decrease the mass to zero.

If we set all $\nu_j = i 3/2 - i 2 \alpha$ and $D=3$ in \eqref{eq:phi2phi2firstorder2}, and examine the $\Gamma$ functions, we will find {\it four} pairs of nearly overlapping zeroth poles which can lead to a maximum of three simultaneous contour pinches\footnote{A very similar phenomenon happens with the massless two-loop calculation in \secref{sec:MasslessMixing}. The divergence structure of the MB integral is explained in greater detail there.}.
That means $\langle \phi^2 \phi^2 \rangle_\lambda$ diverges as $\sim \alpha^{-3}$ where $2\alpha$ is the width of each pinch.
Of these, one factor of $\alpha^{-1}$ is already present at the tree level (cf.\ \eqref{eq:phi2phi2treegeneral2} and \cite{Rajaraman:2010xd}) and another factor of $\alpha^{-1}$ comes from the loop integral, just like in \eqref{eq:Anom1stOrderResult}. The remaining $\alpha^{-1}$ is roughly due to the time evolution of the long wavelength modes; this intuition is made precise by the effective field theory treatment of such modes in \cite{Cohen:2020php,Cohen:2021fzf}.
Such divergences are a well-known feature of massless fields in dS and their careful resummation leads to Starobinsky's Stochastic Inflation framework \cite{Starobinsky:1994bd,Baumgart:2019clc}.


It was realized in a follow up work\footnote{This insight is due to Manuel Loparco.} that the present method can give closed form expressions for the anomalous dimensions by analytic continuation of $\nu$. The details of this method will appear in a future paper, but we quote the result here:
\begin{equation}
  \gamma_{\phi^2} =
    - \frac{
      \cosh(2 \pi \nu)
      \Gamma \left( \frac{3}{2} + i \nu \right)^2
      \Gamma \left( \frac{3}{2} + 2 i \nu \right)
      \Gamma (-i \nu)^2 \lambda
    }{
      16 \pi^{7/2} \Gamma(3 + 2 i \nu)
    } .
    \label{eq:AnomDimClosedForm}
\end{equation}

\section{Mixing of \texorpdfstring{$\phi^3$}{phi cubed} and \texorpdfstring{$\phi$}{phi}}
\label{sec:phi3phiMixing}

As our second example, we compute the order $\lambda$ contribution to $\langle \phi^3 \phi \rangle$.
The dynamics of a massless scalar, as described by Stochastic Inflation, receives an NNLO correction from this term \cite{Cohen:2021fzf}. It is easiest to see this in the Soft de Sitter Effective Theory (SdSET), wherein Stochastic Inflation is a direct consequence of EFT power counting (see sec.\ 5.2 of \cite{Cohen:2020php}). The calculation below produces the same divergence structure for $\langle \phi^3 \phi \rangle_\lambda$ in the full theory as that computed in the SdSET, verifying the correctness of the effective theory approach.

\begin{figure}
  \centering
  \begin{tikzpicture}

      \begin{scope}[xshift=-0.7cm]
        \draw[thick] (-0.5,4) -- (4.5,4) node[right,yshift=-3] {$\tau_0$};
        \draw[thick] (0,4) -- (2,2) -- (4,4);
        \draw[thick] (1.5,4) -- (2,2) -- (2.5,4);
        \filldraw (2,2) circle (2pt) node[below] {$\tau$};

        \filldraw (0,4) circle (1pt) node[above] {$\p_1$};
        \filldraw (1.5,4) circle (1pt) node[anchor=south east] {$\p_2$};
        \filldraw (2.5,4) circle (1pt);
        \node[above] at (2.7,4) {$\k-\p_1-\p_2$};
        \filldraw (4,4) circle (1pt) node[anchor=south west] {$\k$};
      \end{scope}

      \draw[thick] (8,3) circle [radius=1];
      \draw[thick] (7,3) -- (10,3);
      \draw[thick] (9,3) -- (10.5,3) node[right] {$\phi(\k',\tau_0)$};
      \filldraw (9,3) circle (2pt) node[anchor=north west] {$\tau$};

      \filldraw[thick,draw=black,fill=white] (7,3) circle (4pt) node[left] {$[\phi^3](\k,\tau_0)$ \ };
      \node[cross out, draw=black, very thick, scale=0.7] at (7,3) {};
      \node[above] at (8,4) {$\p_1$};
      \node[below] at (8,3) {$\p_2$};
      \node[below] at (8,2) {$\k-\p_1-\p_2$};

      \draw[->, very thick] (3.5,3) -- (4.5,3);

      \draw[->,thick] (-1,2.25) -- (-1,3.5);
      \node[below] at (-1,2.25) {time};
  \end{tikzpicture}
  \caption{\label{fig:phi3phicorr} The leading contribution to $\langle \phi^3 \phi \rangle$.}
\end{figure}
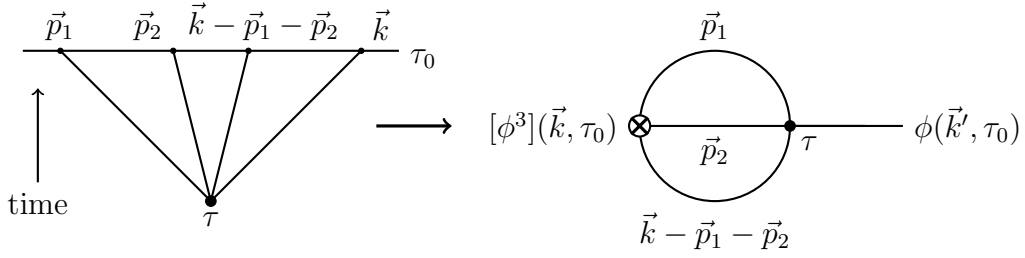

We start with \eqref{eq:Mellin4pt3} and contract three of the legs together resulting in the two loop diagram of \figref{fig:phi3phicorr},
\begin{align}
  \langle \phi^3 \phi \rangle^\prime_\lambda
  &= \int \frac{d^\D p_1}{(2 \pi)^\D} \frac{d^\D p_2}{(2 \pi)^\D}
  \left\langle \phi_{\p_1}^{\left(\nu_{1}\right)} \phi_{\p_2}^{\left(\nu_{2}\right)} \phi_{\k - \p_1 - \p_2}^{\left(\nu_{3}\right)}
  \phi_{\k}^{\left(\nu_{4}\right)} \right\rangle^{\prime} \nonumber \\
  &=
  C \int [ds]_4 \, 2\pi i \, \delta \left( {\ss \frac{\D}{2}} - (s_1+s_2+s_3+s_4) \right) \rho(\bm{s},\bm{\nu}) 2^\D I_{\rm sun} (k,\bm{s},\bm{\nu})
  \label{eq:phi3phiintegral1}
\end{align}
where $C$ is the prefactor defined in \eqref{eq:4ptprefactor} and $I_{\rm sun}$ is the momentum integral
\begin{equation}
  I_{\rm sun} (k,\bm{s},\bm{\nu}) = \int \frac{d^\D p_1}{(2 \pi)^3} \frac{d^\D p_2}{(2 \pi)^3}
      \frac{1}{p_1^{a} \, p_2^{b} \, (\k - \p_1 - \p_2)^{c} \, k^{d}},
  \label{eq:Isun}
\end{equation}
with $a = 2 s_1 - i \nu_1, b = 2 s_2 - i \nu_2, c = 2 s_3 - i \nu_3$ and $d = 2 s_4 - i \nu_4$. Once again, this integral can be evaluated by noting that it is a convolution in momentum space (cf.\ \eqref{eq:ImomConvolution}). We'll also take this opportunity to introduce an analytic continuation parameter $\D - \bar{\D}$, just as we did in \eqref{eq:Imom}. The result is
\begin{align}
  I_{\rm sun} (k,\bm{s},\bm{\nu})
  =
  &\frac{1}{(4 \pi)^{\frac{3\D-\bar{\D}}{2}}}
  \frac{(-k \tau_0)^{\D-\bar{\D}}}{k^{\sum_{j=1}^3 (2 s_j - i \nu_j) - 2\D}} \nonumber \\
    &\quad \prod_{j=1}^{3}
    \frac{\Gamma \left( \frac{\D}{2} - s_j + \frac{i\nu_j}{2} \right)
      }{
        \Gamma \left(s_j - \frac{i\nu_j}{2} \right)
      }
  \frac{
    \Gamma \left( s_1 + s_2 + s_3 - i(\nu_1 + \nu_2 + \nu_3)/2 - \frac{3\D-\bar{\D}}{2} \right)
  }{
    \Gamma \left( \frac{3\D}{2} - s_1 - s_2 - s_3 + i(\nu_1 + \nu_2 + \nu_3)/2 \right)
  } . \label{eq:sunsetintegral2}
\end{align}
%
We can substitute this into \eqref{eq:phi3phiintegral1} and apply the delta function. In order to proceed we must plug in some actual numbers for the masses and dimension.

\subsection{The massless case}
\label{sec:MasslessMixing}

We'll now restrict our attention to the case of massless scalar fields. It is not possible to make \eqref{eq:phi3phiintegral1} well-defined with a single parameter. So we will do the calculation in two different parameterizations, first by floating the masses and then by tweaking the time integral and number of dimensions, and compare the results.
We set
\begin{equation}
  \begin{aligned}
    \nu_j &= i \frac{3}{2} - i 2 \alpha, \quad j=1,2,3 \\
    \nu_4 &= i \frac{3}{2} - i 2 \alpha_4 \\
    D &= 3 \\
    \bar{D} &= 3 + 2 \delta
  \end{aligned}
\end{equation}
where we have introduced three parameters to satisfy \eqref{eq:MellinContourPrescription}. We could have chosen a different $\alpha_j$ for each $\nu_j$, to get $\nu_j = i \frac{3}{2} - i 2 \alpha_j$.
However, these parameters will always show up together as the sum $\alpha_1+\alpha_2+\alpha_3$, which makes sense since the $s_j$ variables associated with the legs carrying loop momenta are interchangeable.
So $\alpha_1+\alpha_2+\alpha_3$ is really just one parameter which we have identified as $3\alpha$ and distributed equally between $\nu_1, \nu_2$ and $\nu_3$. The fourth leg, which does not participate in the loop integral, is given its own $\alpha_4$. We end up with
\begin{equation}
  \langle \phi^3 \phi \rangle^\prime_{\lambda}
  =
  C \frac{8}{(4 \pi)^{3-\delta}}
  \frac{(-k \tau_0)^{-2 \delta}}{k^{3 - 6\alpha - 2\alpha_4}}
  \int [\d s]_3 \, \frac{\Gamma_1 \Gamma_2  \Gamma_3 \Gamma_4 \Gamma_5 \Gamma_6 \Gamma_7 \Gamma_8 \Gamma_9}{\Gamma_{10}},
  \label{eq:phi3phiintegral4}
\end{equation}
where
\begin{equation}
  \begin{aligned}
    \Gamma_1 &= \Gamma( {\ss \frac{9}{4}} - s_1 - s_2 - s_3 - \alpha_4 ) \\
    \Gamma_2 &= \Gamma( {\ss \frac{3}{4}} - s_1 - s_2 - s_3 + \alpha_4 ) \\
    \Gamma_3 &= \Gamma(-{\ss \frac{3}{4}} + s_1 + s_2 + s_3 - 3\alpha + \delta ) \\
    \Gamma_4 &= \Gamma( {\ss \frac{3}{4}} - s_1 + \alpha ) \\
    \Gamma_5 &= \Gamma(-{\ss \frac{3}{4}} + s_1 + \alpha ) \\
    \Gamma_6 &= \Gamma( {\ss \frac{3}{4}} - s_2 + \alpha ) \\
    \Gamma_7 &= \Gamma(-{\ss \frac{3}{4}} + s_2 + \alpha ) \\
    \Gamma_8 &= \Gamma( {\ss \frac{3}{4}} - s_3 + \alpha ) \\
    \Gamma_9 &= \Gamma(-{\ss \frac{3}{4}} + s_3 + \alpha ) \\
    \Gamma_{10} &= \Gamma({\ss \frac{9}{4}} - s_1 - s_2 - s_3 + 3\alpha ) .
  \end{aligned}
  \label{eq:SunsetGammas}
\end{equation}
We will now focus on evaluating the MB integral in \eqref{eq:phi3phiintegral4},
\begin{equation}
  K = \int [\d s]_3 \, \frac{\Gamma_1 \Gamma_2  \Gamma_3 \Gamma_4 \Gamma_5 \Gamma_6 \Gamma_7 \Gamma_8 \Gamma_9}{\Gamma_{10}}. \label{eq:SunsetK}
\end{equation}

\subsubsection{Anticipating the answer}
\label{sec:DivergenceStructure}

At a glance \eqref{eq:SunsetK} looks very similar to \eqref{eq:AnomK} but it is in fact hiding a much more intricate divergence structure. This is already apparent if we examine the arguments of the $\Gamma$ functions above. For instance, the zeroth left pole of $\Gamma_4$ is at a distance $2\alpha$ from the zeroth right pole of $\Gamma_5$. As we take $\alpha \to 0$ these poles pinch the contour and generate a $1/2\alpha$ divergence. There are four such pairs of $\Gamma$ functions in the list which together can generate up to a cubic order divergence.
Thus, without actually evaluating \eqref{eq:SunsetK}, we may deduce the following form of the answer:
\begin{equation}
  K \sim \frac{c_3}{{\bm \alpha}^3} + \frac{c_2}{{\bm \alpha}^2} + \frac{c_1}{{\bm \alpha}} + O\left({\bm \alpha}^0\right) ,
  \label{eq:DivergenceStructure}
\end{equation}
where ${\bm \alpha}$ denotes some linear combination of the parameters $\alpha, \alpha_4$ and $\delta$, all of which are nearly zero, and $c_i$ collects together the remaining factors. Also note that, for a massive scalar ($|\nu| < 3/2$) there would be fewer simultaneous contour pinches, and the leading small ${\bm \alpha}$ behavior would be less singular than \eqref{eq:DivergenceStructure}.
This is another example of our observation from \figref{fig:AnomDimensionsVsMasses}: massless scalars in dS are saddled with more IR divergences than their massive counterparts.

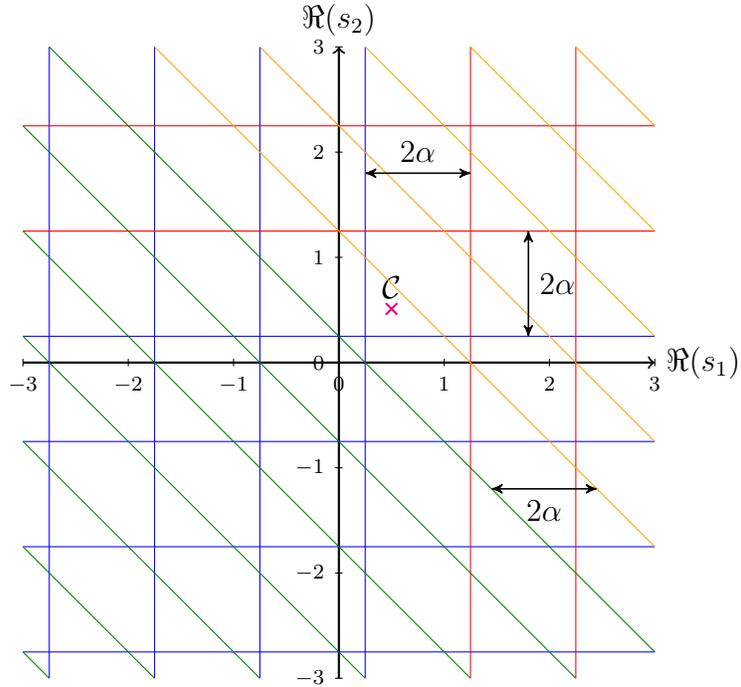
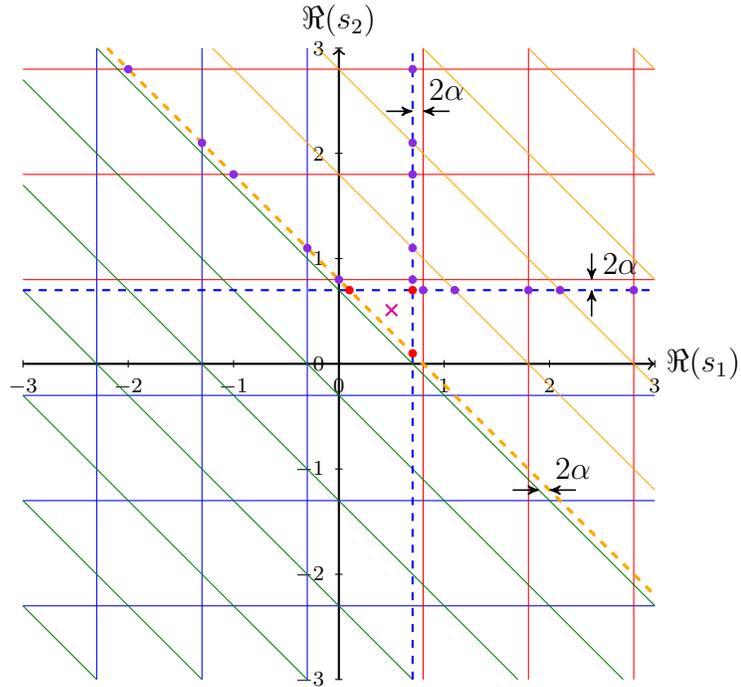
\begin{figure}
  \begin{subfigure}{\textwidth}
  \centering
  \begin{tikzpicture}[scale=1.4]
  {
    \def \prpp {5}
    \def \prp {4}
    \def \pr {3} 
    \def \prm {2}
    \def \prmm {1}
    \draw[thick,->] (-\pr,0) -- (\pr,0) node[right] {$\Re(s_1)$};
    \draw[thick,->] (0,-\pr) -- (0,\pr) node[above] {$\Re(s_2)$};
    \foreach \x in {-\pr,...,\pr}
      \draw (\x cm,1pt) -- (\x cm,-1pt) node[anchor=north] {$\ss \x$};
    \foreach \y in {-\pr,...,\pr}
        \draw (1pt,\y cm) -- (-1pt,\y cm) node[anchor=east] {$\ss \y$};

    \node[cross out, draw=magenta, thick, scale=0.5] at ({0.5},{0.51}) {};
    \node[above] at ({0.5},{0.51}) {$\C$};

    {
    \def \a {0.5}
    \foreach \n in {0,...,\pr}
      \draw[draw=blue] ({0.75-\a-\n},-\pr) -- ({0.75-\a-\n},\pr);
    \foreach \n in {0,...,\pr}
      \draw[draw=blue] (-\pr,{0.75-\a-\n}) -- (\pr,{0.75-\a-\n});

    \foreach \n in {0,...,\prmm}
      \draw[draw=red] ({0.75+\a+\n},-\pr) -- ({0.75+\a+\n},\pr);
    \foreach \n in {0,...,\prmm}
      \draw[draw=red] (-\pr,{0.75+\a+\n}) -- (\pr,{0.75+\a+\n});

    \foreach \n in {1,...,\prpp}
      \draw[draw=Green] (-\pr,{0.75-\a-\n+\pr}) -- ({0.75-\a-\n+\pr},-\pr);
    \draw[draw=Green] ({0.75-\a-\pr},\pr) -- (\pr,{0.75-\a-\pr});
    \draw[draw=Green] (-\pr,{0.75-\a-3}) -- ({0.75-\a-3},-\pr);

    \foreach \n in {0,...,\prp}
      \draw[draw=Orange] ({0.75+\a+\n-\pr},\pr) -- (\pr,{0.75+\a+\n-\pr});

    \draw[semithick,{stealth'}-{stealth'}] ({0.75-\a},{3*\pr/5}) -- ({0.75+\a},{3*\pr/5}) node[above,pos=0.5] {$2 \alpha$};
    \draw[semithick,{stealth'}-{stealth'}] ({3*\pr/5},{0.75-\a}) -- ({3*\pr/5},{0.75+\a}) node[right,pos=0.5] {$2 \alpha$};
    \draw[semithick,{stealth'}-{stealth'}] ({0.75-\a+2*\pr/5},{-2*\pr/5}) -- ({0.75+\a+2*\pr/5},{-2*\pr/5}) node[below,pos=0.5] {$2 \alpha$};
    }
  }
  \end{tikzpicture}
  \caption{\label{fig:BeforeAC} Prior to analytic continuation, when the left/right poles are separated (cf.\ \figref{fig:PolesSeparated}).}
  \end{subfigure}
  %
  %
  \begin{subfigure}{\textwidth}
  \centering
  \begin{tikzpicture}[scale=1.4]
  {
    \def \prppp {6}
    \def \prpp {5}
    \def \prp {4}
    \def \pr {3} 
    \def \prm {2}
    \def \prmm {1}
    \draw[thick,->] (-\pr,0) -- (\pr,0) node[right] {$\Re(s_1)$};
    \draw[thick,->] (0,-\pr) -- (0,\pr) node[above] {$\Re(s_2)$};
    \foreach \x in {-\pr,...,\pr}
      \draw (\x cm,1pt) -- (\x cm,-1pt) node[anchor=north] {$\ss \x$};
    \foreach \y in {-\pr,...,\pr}
        \draw (1pt,\y cm) -- (-1pt,\y cm) node[anchor=east] {$\ss \y$};

    \node[cross out, draw=magenta, thick, scale=0.5] at ({0.5},{0.51}) {};

    {
    \def \a {0.05}
    \foreach \n in {1,...,\pr}
      \draw[draw=blue] ({0.75-\a-\n},-\pr) -- ({0.75-\a-\n},\pr);
    \foreach \n in {1,...,\pr}
      \draw[draw=blue] (-\pr,{0.75-\a-\n}) -- (\pr,{0.75-\a-\n});

    \foreach \n in {0,...,\prm}
      \draw[draw=red] ({0.75+\a+\n},-\pr) -- ({0.75+\a+\n},\pr);
    \foreach \n in {0,...,\prm}
      \draw[draw=red] (-\pr,{0.75+\a+\n}) -- (\pr,{0.75+\a+\n});

    \foreach \n in {1,...,\prppp}
      \draw[draw=Green] (-\pr,{0.75-\a-\n+\pr}) -- ({0.75-\a-\n+\pr},-\pr);
    \draw[draw=Green] ({0.75-\a-\pr},\pr) -- (\pr,{0.75-\a-\pr});

    \foreach \n in {1,...,\prpp}
      \draw[draw=Orange] ({0.75+\a+\n-\pr},\pr) -- (\pr,{0.75+\a+\n-\pr});

    \draw[thick,dashed,draw=blue] ({0.75-\a},-\pr) -- ({0.75-\a},\pr);
    \draw[thick,dashed,draw=blue] (-\pr,{0.75-\a}) -- (\pr,{0.75-\a});
    \draw[very thick,dashed,draw=Orange] ({0.75+\a-\pr},\pr) -- (\pr,{0.75+\a-\pr});

    \def \pca {Red}
    \filldraw[draw=\pca,fill=\pca] ({0.75-\a},{0.75-\a}) circle (1pt) node[above] {};
    \filldraw[draw=\pca,fill=\pca] ({2*\a},{0.75-\a}) circle (1pt) node[above] {};
    \filldraw[draw=\pca,fill=\pca] ({0.75-\a},{2*\a}) circle (1pt) node[above] {};

    \def \pcb {BlueViolet}
    \foreach \n in {0,...,\prm}
      \filldraw[draw=\pcb,fill=\pcb] ({0.75-\a},{0.75+\a+\n}) circle (1pt) node[above] {};
    \foreach \n in {0,...,\prm}
      \filldraw[draw=\pcb,fill=\pcb] ({0.75+\a+\n},{0.75-\a}) circle (1pt) node[above] {};
    \foreach \n in {1,...,\prm}
      \filldraw[draw=\pcb,fill=\pcb] ({\n+2*\a},{0.75-\a}) circle (1pt) node[above] {};
    \foreach \n in {1,...,\prm}
      \filldraw[draw=\pcb,fill=\pcb] ({0.75-\a},{\n+2*\a}) circle (1pt) node[above] {};
    \foreach \n in {0,...,\prm}
      \filldraw[draw=\pcb,fill=\pcb] ({-\n},{0.75+\a+\n}) circle (1pt) node[above] {};
    \foreach \n in {1,...,\prm}
      \filldraw[draw=\pcb,fill=\pcb] ({0.75-\a-\n},{\n+2*\a}) circle (1pt) node[above] {};

    \draw[semithick,-{stealth'}] ({0.75-\a-0.25},{4*\pr/5}) -- ({0.75-\a},{4*\pr/5});
    \draw[semithick,-{stealth'}] ({0.75+\a+0.25},{4*\pr/5}) -- ({0.75+\a},{4*\pr/5});
    \node[anchor=south west] at (0.75,4*\pr/5) {$2 \alpha$};
    \draw[semithick,-{stealth'}] ({4*\pr/5},{0.75-\a-0.25}) -- ({4*\pr/5},{0.75-\a});
    \draw[semithick,-{stealth'}] ({4*\pr/5},{0.75+\a+0.25}) -- ({4*\pr/5},{0.75+\a});
    \node[anchor=south west] at (4*\pr/5,0.75) {$2 \alpha$};
    \draw[semithick,-{stealth'}] ({0.75-\a+2*\pr/5-0.25},{-2*\pr/5}) -- ({0.75-\a+2*\pr/5},{-2*\pr/5});
    \draw[semithick,-{stealth'}] ({0.75+\a+2*\pr/5+0.25},{-2*\pr/5}) -- ({0.75+\a+2*\pr/5},{-2*\pr/5});
    \node[anchor=south west] at ({0.75+2*\pr/5},{-2*\pr/5}) {$2 \alpha$};
    }
  }
  \end{tikzpicture}
  \caption{\label{fig:AfterAC} At the end of continuation, with the dashed lines indicating crossed poles (cf.\ \figref{fig:PolesCrossed}).}
  \end{subfigure}
  \caption{\label{fig:SingularStructure} The poles of the integrand are represented by straight lines on a $\Re(s_1)-\Re(s_2)$ plane. Residues are computed at the intersections of these lines. The ${\bm \times}$ symbol in (a) represents the real part of the $s_1$ and $s_2$ contours. Poles that cross over to the other side are represented by dashed lines in (b).
  The residues at {\textcolor{Red}{$\bullet$}} contain a $\Gamma(\alpha)^2$ divergence whereas those at {\textcolor{BlueViolet}{$\bullet$}} give a $\Gamma(\alpha)$. The latter encode subhorizon physics, as explained in \secref{sec:DivergenceStructure}.
  }
\end{figure}

The Mellin variables $s_j$ carry physical meaning, as eigenvalues of the dilatation operator. Therefore, it is worth understanding which poles contribute to the terms in \eqref{eq:DivergenceStructure}. To do so we consider a simpler 2-fold toy integral, with the same features as \eqref{eq:SunsetK}:
\begin{equation}
  K_{\rm toy} = \int [\d s]_2 \Gamma( {\ss \frac{3}{4}} - s_1 - s_2 + \alpha ) \Gamma( -{\ss \frac{3}{4}} + s_1 + s_2 + \alpha ) \prod_{j=1}^2 \Gamma( {\ss \frac{3}{4}} - s_j + \alpha ) \Gamma( -{\ss \frac{3}{4}} + s_j + \alpha ) .
\end{equation}
As $\alpha \to 0$ this integrand generates three contour pinches, of which at most two manifest simultaneously. Thus, the answer has the form
\begin{equation}
  K_{\rm toy} \sim \frac{\bar{c}_2}{\alpha^2} + \frac{\bar{c}_1}{\alpha} + O(\alpha^0).
  \label{eq:ToyKAnswer}
\end{equation}
We can examine the singular structure of $K_{\rm toy}$ on a $\Re(s_1)-\Re(s_2)$ plane (\figref{fig:SingularStructure}).
The poles of our integrand are real and these are represented by straight lines in \figref{fig:BeforeAC} and \figref{fig:AfterAC}. These lines are the graphs $U_i({\bm s}) = -n$ (cf.\ \eqref{eq:MBPoles}).
For example, the red lines represent the poles of $\Gamma( 3/4 - s_j + \alpha )$, which are at $s_{j\star} = 3/4 + \alpha + n$ with $n \in \mathbb{Z}_0$ and $j=1,2$.
Similarly, we map:
\begin{align*}
  \frac{3}{4} - s_j + \alpha &= -n \text{ (\textcolor{red}{red} lines)} \\
  -\frac{3}{4} + s_j + \alpha &= -n \text{ (\textcolor{blue}{blue} lines)} \\
  \frac{3}{4} - s_1 - s_2 + \alpha &= -n \text{ (\textcolor{orange}{orange} lines)} \\
  -\frac{3}{4} + s_1 + s_2 + \alpha &= -n \text{ (\textcolor{Green}{green} lines)}.
\end{align*}
In applying Cauchy's theorem we compute residues at the intersections of these lines and add them up in specific ways \cite{Friot:2011ic}. The real parts of the straight line contours are represented by a point $\C$ in these plots.

We define the integral by choosing a value of $\alpha$ such that the left/right poles are well separated by contours, as shown in \figref{fig:BeforeAC} (see also \figref{fig:PolesSeparated}). As we decrease $\alpha \to 0$ some of the $n=0$ poles (blue and orange) cross over to the other side, and come within $2\alpha$ distance of $n=0$ poles of the opposite nature (red and green lines).
The crossed poles are indicated by dashed lines in \figref{fig:AfterAC}, and the intersections at which we take residues are marked with dots. Due to their proximity to other poles, the residues along each dashed line will contain a factor of $\Gamma(2\alpha)$ (cf.\ \eqref{eq:DivergentResidue}); residues computed at intersections of two dashed lines generate $\Gamma(2\alpha)^2$. Residues computed at all other intersections, which don't involve any dashed lines, contribute to the $O(\alpha^0)$ term in \eqref{eq:ToyKAnswer}.

It is clear from \figref{fig:SingularStructure} that only a small subset of all possible poles contribute divergences to the answer. This is just a 2-dimensional generalization of what we observed in \eqref{eq:SeparatingADivergence1}. However,  there is something else going on: only a {\it finite} number of intersections produce an $\alpha^{-2}$ term, whereas an infinity of them diverge as $\alpha^{-1}$.
The poles that generate the $\alpha^{-2}$ capture the leading behavior of the mode functions in the long wavelength limit (see discussion under \eqref{eq:LateTime2ptFn}). In fact, the EFT description of such soft modes reproduces this divergence exactly, without any UV matching \cite{Cohen:2021fzf}. On the other hand, an infinite number of residues have to be summed over to fully account for the $\alpha^{-1}$ term, indicating that it is tracking much more than just $k \ll (aH)$ effects. In other words, this term encodes subhorizon physics.
A similar analysis on the example from \secref{sec:AnomDimConformal} would show that the $\alpha^{-1}$ term there is also derived from summing over an infinite number of intersections (this is apparent from \eqref{eq:K39Integral} and \eqref{eq:K108Integral}), which means the anomalous scaling we found there was truly a UV effect.

Diagrams like \figref{fig:SingularStructure} have special significance in the evaluation of multidimensional MB integrals by the method of residues \cite{Zhdanov1998}. The straight lines we studied above become hypersurfaces for an $N$-fold MB integral, and we take residues over the polyhedra formed by these surfaces. For a field theory calculation, this leads to a multiple series in powers and logarithms of the kinematic parameters. The convergence of such a series has a geometrical connection to the aforementioned polyhedra \cite{Friot:2011ic,Ananthanarayan:2020fhl}. While we focus only on the divergent contributions in this work, it would be worth investigating whether the ideas in these references can be used to understand the properties of dS loops in various kinematic limits.


%

\subsubsection{Choosing the contours}

We now return to the task of computing \eqref{eq:SunsetK}. The intial values of the parameters $\alpha, \alpha_4$ and $\delta$ are chosen to satisfy \eqref{eq:MellinContourPrescription}, by the same process as in \secref{sec:ChoosingContoursAnom}.
One possible choice is $\delta = 0.06, \alpha = 0.26, \alpha_4 = 0.76$ with contours $s_1^\C = s_2^\C = s_3^\C = 0.495$. We have placed the contours at the same position in the $s_1, s_2$, and $s_3$ planes to leverage the symmetry of the integral under the exchange of these variables. In practice, this choice leads to simultaneous pole crossings, which requires special care \cite{Anastasiou:2005cb}. A simple solution that works for the present calculation is to stagger the contours by a little bit, by choosing say $s_1^\C=0.5, s_2^\C=0.495$, and $s_3^\C=0.492$.

The conditions $U_{4,5}(\C)>0, U_{6,7}(\C)>0, U_{8,9}(\C)>0$, $U_{2,3}(\C)>0$, and $U_{1,5,7,9}(\C)>0$ produce pinches similar to \eqref{eq:AnomConstraintCombined2} and \eqref{eq:AnomConstraintCombined3},
\begin{equation}
  \begin{aligned}
    \frac{3}{4} - \alpha &< s^{\C}_j < \frac{3}{4} + \alpha, \quad  j=1,2,3 \\
    \frac{3}{4} + 3\alpha - \delta &< s_1^\C + s_2^\C + s_3^\C < \frac{3}{4} + \alpha_4 \\
    \frac{9}{4} - 3\alpha &< s_1^\C + s_2^\C + s_3^\C < \frac{9}{4} - \alpha_4
    \label{eq:SunsetNotableConstraints}
  \end{aligned}
\end{equation}
from which we infer that $\alpha>0$, $3 \alpha - \delta < \alpha_4$ and $3\alpha > \alpha_4$. We should honor these relationships throughout, lest we end up with incorrect signs for the divergences. It should be noted that the last pinch above does not give a divergence, because the poles of $\Gamma_1$ are very nearly the same as those of the $\Gamma_{10}$ in the denominator, and they cancel.
In other words, it does not correspond to an overlap of poles in the unregulated integral, for which $\alpha, \alpha_4$ and $\delta$ are zero.

\subsubsection{Analytic continuation}

We now begin our journey to $\alpha, \alpha_4, \delta \sim 0$ to determine \eqref{eq:SunsetK} for a massless theory. The first poles to cross at those of $\Gamma_5, \Gamma_7$, and $\Gamma_9$. We separate the residues at those poles and write
\begin{equation}
  K \to K_5 + K_7 + K_9 + K_{57} + K_{59} + K_{79} +K_{579} + K .
  \label{eq:KbreakupL1}
\end{equation}
$\Gamma_5, \Gamma_7$ and $\Gamma_9$ are the same upto an exchange of $s_j$, and so are the terms $K_5, K_7$ and $K_9$ as well as $K_{57}, K_{59}$ and $K_{79}$. Therefore, the above breakup is really $K \to 3 K_5 + 3 K_{57} + K_{579} + K$. We focus now on each of these terms. First, $K_{579}$ is the residue of $K$ at the three poles $s_j=\frac{3}{4}-\alpha$ and it involves no integrals.
Next, $K_{57}$ is a single integral which further breaks up, in stages, into $K_{57} \to K_{572} + K_{572'} + K_{57}$ as the parameters are taken to zero. The prime in $K_{572'}$ indicates that the residue is taken at $n=1$ pole of $\Gamma_2$ (cf.\ \eqref{eq:PoleCrossingCondition}).
At the same time, the double integral $K_5$ further breaks up into $K_5 \to K_{527'} + K_{52} + K_5$. Finally, the $K$ in the r.h.s.\ of \eqref{eq:KbreakupL1} is a triple integral which separates as $K \to K_2 + K$. The analytic continuation is now complete, with
\begin{equation}
  K =
    3 (K_{527'} + K_{52} + K_5) + 3 (K_{572} + K_{572'} + K_{57}) + K_{579} + K_2 + K . \label{eq:SunsetKbreakup}
\end{equation}

\subsubsection{Evaluation of the terms}

We will evaluate a few of the terms in \eqref{eq:SunsetKbreakup} to demonstrate the steps involved. The easiest ones are those involving residues in all three variables. For e.g.
\begin{align}
  K_{579}
  &= \frac{\Gamma_1 \Gamma_2  \Gamma_3 \Gamma_4 \ \ \Gamma_6 \ \ \Gamma_8 \ \ }{\Gamma_{10}} \bigg \rvert_{s_j=\frac{3}{4}-\alpha}
  =
  \frac{
    \Gamma (2 \alpha )^3
    \Gamma (3 \alpha -\alpha_4)
    \Gamma \left( - \frac{3}{2} + 3 \alpha + \alpha_4 \right)
    \Gamma \left( \frac{3}{2} - 6 \alpha + \delta \right)
  }{
    \Gamma (6 \alpha )
  }. \label{eq:K579Final}
\end{align}
There are also terms with a single integral, like
\begin{align}
  K_{52}
  &= \int_{\C_3} \frac{\d s_3}{2 \pi i} \,
  \frac{\Gamma_1 \ \ \Gamma_3 \Gamma_4 \ \ \Gamma_6 \Gamma_7^{**} \, \Gamma_8 \Gamma_9^*}{\Gamma_{10}} \bigg \rvert_{s_1=\frac{3}{4}-\alpha, \, s_2=-s_3+\alpha+\alpha_4} \nonumber \\[1em]
  &=
  \Gamma( 2 \alpha )
  \Gamma( - 3\alpha + \alpha_4 + \delta )
  \frac{
    \Gamma( \frac{3}{2} - 2 \alpha_4 )
  }{
    \Gamma({\ss \frac{3}{2}} + 3\alpha - \alpha_4 )
  } \\
  &\quad \times
  \int_{\C_3} \frac{\d s_3}{2 \pi i} \,
    \Gamma \left( {\ss \frac{3}{4}} + s_3 - \alpha_4 \right)
    \Gamma^* \left( -{\ss \frac{3}{4}} + s_3 + \alpha \right)
    \Gamma \left( {\ss \frac{3}{4}} - s_3 + \alpha \right)
    \Gamma^{**} \left( -{\ss \frac{3}{4}} - s_3 + 2\alpha + \alpha_4 \right) .
    \nonumber
\end{align}
The symbol $\Gamma^{*}$ indicates that the first pole of that $\Gamma$ function has crossed over. $\Gamma^{**}$ means the first two poles have crossed, and so on (see \secref{sec:GeneralAC}). The $s_3$ integral by itself does not produce any new divergence since the overlapping poles are on the other side of the contour (see \figref{fig:IncorrectContour}).
Since the prefactor is already $O(\alpha^{-2})$ as $\alpha \to 0$, we need to retain the $\alpha$ dependence inside the $s_3$ integral to compute the $O(\alpha^{-1})$ contribution to $K$. We can Taylor expand the integrand to linear order in $\alpha, \alpha_4$ and write
\begin{align} \label{eq:TaylorExpandedK52s}
\int_{\C_3} \frac{\d s_3}{2 \pi i} \,
  \Gamma      &\left( {\ss \frac{3}{4}} + s_3 \right)
  \Gamma^*    \left( -{\ss \frac{3}{4}} + s_3 \right)
  \Gamma      \left( {\ss \frac{3}{4}} - s_3 \right)
  \Gamma^{**} \left( -{\ss \frac{3}{4}} - s_3 \right) \\
  &\times \bigg( 1 +
    \alpha \psi(-{\ss \frac{3}{4}} + s_3)
    + \alpha \psi( {\ss \frac{3}{4}} - s_3)
    + (2\alpha + \alpha_4) \psi(-{\ss \frac{3}{4}} - s_3)
    - \alpha_4 \psi({\ss \frac{3}{4}} + s_3)
  \bigg) \nonumber
\end{align}
where $\psi(x) \equiv \frac{\Gamma'(x)}{\Gamma(x)}$ is the digamma function. This integral can be computed using corollaries of Barnes's first lemma \eqref{eq:B1L}, like those given in appendix D of \cite{Smirnov:2004ym}. The answer is
\begingroup
\allowdisplaybreaks[0]
\begin{align}
  &K_{52} =
  \Gamma( 2 \alpha )
  \Gamma( - 3\alpha + \alpha_4 + \delta )
  \frac{
    \Gamma( \frac{3}{2} - 2 \alpha_4 )
  }{
    \Gamma({\ss \frac{3}{2}} + 3\alpha - \alpha_4 )
  }
  \bigg(
  -2 \pi + \frac{4 \pi}{3} [\gamma_E + \psi({\ss \frac{3}{2}})] \nonumber \\[0.75em]
  - &\alpha \frac{2 \pi}{3}
    \bigg[(2 \gamma_E^2
      + \pi^2
      + 9 \psi({\ss -\frac{1}{2}})
      - 3 \psi({\ss \frac{1}{2}}) \nonumber
      - 6 \psi({\ss -\frac{3}{2}}) \psi({\ss \frac{3}{2}})
      + \gamma_E (- 6 - 6\psi({\ss -\frac{3}{2}}) + 2 \psi({\ss \frac{3}{2}}))
    \bigg]
  \\[0.75em]
  - &\alpha_4\frac{2}{3} \pi
  \bigg[
    -8 + \pi^2 -2 \gamma_E
    \psi(-{\ss \frac{3}{2}})+3 \psi(-{\ss \frac{1}{2}})
      -3 \psi({\ss \frac{1}{2}})
     + 2 \gamma_E \psi({\ss \frac{3}{2}})
     - 2 \psi(-{\ss \frac{3}{2}})\psi({\ss\frac{3}{2}})
     + 2 \psi({\ss \frac{3}{2}})^2
   \bigg]
  \bigg)
  \label{eq:K52sFinal}
\end{align}
\endgroup
Next, there are terms in \eqref{eq:SunsetKbreakup} which involve double integrals. For instance,
\begin{equation}
\begin{aligned}
  K_5 &=
    \int_{\C_2} \frac{\d s_2}{2 \pi i}
    \int_{\C_3} \frac{\d s_3}{2 \pi i} \,
    \frac{\Gamma_1 \Gamma_2^*  \Gamma_3 \Gamma_4 \ \ \Gamma_6 \Gamma_7^* \Gamma_8 \Gamma_9^*}{\Gamma_{10}} \bigg \rvert_{s_1=\frac{3}{4}-\alpha} \\[1em]
    &=
    \Gamma(2 \alpha)
    \int_{\C_2} \frac{\d s_2}{2 \pi i}
      \Gamma    ( {\ss \frac{3}{4}} - s_2 )
      \Gamma^*  (-{\ss \frac{3}{4}} + s_2 )
    \int_{\C_3} \frac{\d s_3}{2 \pi i} \,
      \Gamma ( {\ss \frac{3}{4}} - s_3 )
      \Gamma^* ( -{\ss \frac{3}{4}} + s_3 )
      \Gamma^* ( -s_{23} )
      \Gamma ( s_{23} ) \label{eq:K5leadingorder}
\end{aligned}
\end{equation}
where $s_{23} \deq s_2 + s_3$.
We have retained only the leading order contribution from the double integral since the prefactor is at most $O(\alpha^{-1})$. Once again these are evaluated using corollaries to \eqref{eq:B1L}, and we get
\begin{equation}
  K_5 =
  \Gamma(2 \alpha)
  \frac{2 \pi}{3}
    \bigg[3 \gamma_E^2
      + \pi^2
      + 6 \gamma_E \, \psi({\ss \frac{3}{2}})
      + 3 \psi({\ss \frac{3}{2}})^2 - 12
    \bigg] . \label{eq:K5Final}
\end{equation}
Finally, the $K$ in the r.h.s.\ of \eqref{eq:SunsetK} is what is left of the original $K$ after all the divergent residues have been extracted. It produces at most an $O(1)$ contribution and may be ignored. All other terms in \eqref{eq:SunsetKbreakup} can be evaluated in the manner shown above.

\subsubsection{The problem of too many parameters}
\label{sec:TooManyParams}

It is clear from \eqref{eq:K579Final}, \eqref{eq:K52sFinal} and \eqref{eq:K5Final} that $K$ will have the structure we anticipated in \eqref{eq:DivergenceStructure}.
However, the answer involves three parameters $\alpha, \alpha_4$, and $\delta$, and we need to establish a relationship between these to extract a meaningful result from \eqref{eq:SunsetKbreakup}.
So far, all we have are the inequalities $\alpha>0, 3 \alpha - \delta < \alpha_4$ and $3\alpha > \alpha_4$ (cf.\ \eqref{eq:SunsetNotableConstraints}).
We encountered a similar problem in \eqref{eq:AnomDimTwoParamResult} and resolved it by insisting that all pinches have the same width. Said differently, all divergent $\Gamma$ functions in $K$ must have the same argument. In the present case, such a requirement furnishes the condition (cf.\ \eqref{eq:K52sFinal})
\begin{equation}
  -3 \alpha + \alpha_4 + \delta = 2 \alpha .
  \label{eq:SunsetPinchConstraint1}
\end{equation}
Next, we see that \eqref{eq:K579Final} has a $\Gamma(3\alpha - \alpha_4)$ in the numerator but, as noted under \eqref{eq:SunsetNotableConstraints}, this does not lead to a divergence due to the $\Gamma(6\alpha)$ in the denominator. So it makes little sense to equate $3\alpha - \alpha_4$ with the pinch width $2 \alpha$.
Thus we are still one constraint shy of being able to express $\delta$ and $\alpha_4$ in terms of $\alpha$. To proceed we'll make yet another `reasonable' assumption: we'll choose $\alpha_4 = -3 \alpha$ to make $\Gamma(3\alpha - \alpha_4)$ cancel $\Gamma(6\alpha)$ exactly. Then,
%
\begin{eBox}
  \begin{equation}
  \begin{aligned}
    \langle \phi^3 \phi \rangle^\prime_\lambda \approx
      \frac{H^4}{512 \pi^5}
      \frac{(-k \tau_0)^{-16 \alpha}}{k^3}
      \bigg[
        \frac{\pi}{3 \alpha^3}
        &- \frac{8 \pi (\gamma_E - 2 +\log (4))}{3 \alpha ^2} \\[0.75em]
        &+ \frac{4 \pi \left(3 \pi^2 - 32 + 8 (\gamma_E - 2 + \log (4))^2 \right)}{3 \alpha}
        + O(\alpha^0)
      \bigg]
  \end{aligned} \label{eq:phi3phiFinal}
  \end{equation}
\end{eBox}
where we have used \eqref{eq:SunsetPinchConstraint1} to eliminate $\delta$.


\subsubsection{An alternate parameterization}
\label{sec:SunsetAlternateParam}


We will now redo the calculation using a different parameterization and compare with the above result. This will help us justify our choice of $\alpha_4$ at the end of the previous section.

It is possible to make \eqref{eq:SunsetK} well-defined with just two parameters instead of three. The first of these parameters, $\e$, is introduced by modifying the time integral as in \eqref{eq:TimeIntegralWithParameter}. The second parameter $\kappa$ is a shift in the number of space dimensions, $\D=3+2\kappa$.
Finally we set all $\nu_j \to i \frac{3}{2}$, and $\bar{\D}=3$ in \eqref{eq:sunsetintegral2}.
%
Applying the conditions \eqref{eq:MellinContourPrescription} on this new integral we identify the pinches
\begin{equation}
  \begin{aligned}
    \frac{3}{4} &< s^\C_j < \frac{3}{4} + \kappa; \quad j=1,2,3 \ \\
    \frac{3}{4} + 2\kappa &< s_1^\C + s_2^\C + s_3^\C < \frac{3}{4} + \kappa + \e \\
    \frac{9}{4} &< s_1^\C + s_2^\C + s_3^\C < \frac{9}{4} + \kappa + \e .
    \label{eq:SunsetAlternateConstraints}
  \end{aligned}
\end{equation}
Comparing this with \eqref{eq:SunsetNotableConstraints} we see that the width of the first three pinches is now $\kappa$ instead of $2 \alpha$, the second pinch is $-\kappa + \e$ wide, and the parameters must satisfy $\e > \kappa > 0$.
The last inequality in \eqref{eq:SunsetAlternateConstraints} does not represent a true pinch because the divergence associated with it is canceled by the $\Gamma_{10}$ in the denominator, exactly as in the previous parameterization. However, the important difference is that equating the pinch widths establishes an unambiguous relationship, $\e = 2 \kappa$, between all two parameters.
This also sets $\Gamma_1/\Gamma_{10} \to 1$, which is exactly what the choice $\alpha_4 = -3 \alpha$ accomplished in the last calculation. Proceeding with the analytic continutation, and setting $\e = 2 \kappa$, we arrive at
\begin{equation}
\begin{aligned}
  \langle \phi^3 \phi \rangle^\prime_\lambda \approx
    &\frac{H^{4} (-H \tau_0)^{4 \kappa}}{512 \pi^5}
    \frac{(-k \tau_0)^{-2 \kappa}}{k^{3+2\kappa}}
    \bigg[
      \frac{8 \pi}{3 \kappa^3}
      - \frac{16 \pi (2 \gamma_E - 5 + 2\log(4))}{3 \kappa^2} \\[0.75em]
      &\quad + \frac{4\pi  \left(40 + 3 \pi^2 - 16 (\gamma_E - 2 + \log(4)) + 16 (\gamma_E - 2 + \log (4))^2\right)}{3\kappa}
      + O\left(\kappa^0\right)
    \bigg]
\end{aligned} \label{eq:phi3phiAlternate}
\end{equation}
The leading coefficient of $\log(-k \tau_0)$ is the same in both \eqref{eq:phi3phiFinal} and \eqref{eq:phi3phiAlternate}, upto a renaming of $\alpha \to \kappa$,
\begin{equation}
  - \frac{H^4}{96 \pi^4 k^3 \alpha^2} .
\end{equation}
A more careful renormalization is required to make the subleading terms match. The important takeaway is that the divergence structure of $\langle \phi^3 \phi \rangle^\prime_\lambda$ in the full theory matches what we calculated in the SdSET \cite{Cohen:2021fzf}, with a regulator that also characterizes divergences with $1/\alpha$ poles. Thus, the calculation in this section confirms that the SdSET correctly reproduces the IR divergences of the full theory.


In closing, we note that the sample calculations in this paper were `simple', in that we went no farther than $O(\lambda)$, and the MB integrands did not contain terms of the form $p_i^{s_j}$, where $p_i$ denotes a momentum variable. While it is certainly worth pushing the method to do higher loops, we believe the two examples considered above sufficiently illustrate the usefulness of the method both as a computational tool and as a way to glean qualitative insights about the IR behavior of fields in dS.

\section{Conclusions}
\label{sec:Conclusion}

In this paper, we advanced a method to identify and extract divergences in dS loop calculations. Motivated by the dilatation invariance of dS, we worked with the Mellin-Barnes representation of dS correlation functions \cite{Sleight:2019mgd,Sleight:2019hfp}. This allowed us to import the techniques developed in \cite{Boos:1990rg,Smirnov:2009up, Tausk:1999vh, Anastasiou:2005cb, Czakon:2005rk}, for flat space Feynman diagrams, to the calculation of loop integrals in dS. The resulting expressions have the familiar structure of a dimreg answer (cf.\ \eqref{eq:Anom1stOrderResult} and \eqref{eq:phi3phiFinal}), and we can resum the divergences with dynamical RG by way of extracting meaningful physics. While the examples we considered in this work were simple, they illustrate a few important aspects of loop calculations in dS that are worth highlighting once again:

\begin{enumerate}[wide, labelwidth=!]
  \item The divergences in dS correlation functions manifest as pole overlaps in the Mellin space; the problem of isolating divergences becomes a matter of locating these pole overlaps in a hyperspace spanned by the complex Mellin variables $\{s_\ell\}$.
  \item Our method works for loop integrals involving scalar fields of any mass. In particular, the calculations of \secref{sec:AnomDim} reveal that {\it all} massive scalars are afflicted with secular growth, with additional divergences introduced as we approach the massless limit (see \figref{fig:AnomDimensionsVsMasses}). Stochastic Inflation provides a non-perturbative resolution of these IR issues for massless scalars. The technique developed in this paper allows us to compute higher order corrections to the stochastic description, directly from the full theory.
  %
  \item Dynamical RG resums the secular logs from dS loop integrals, just as regular RG treats UV logs in flat space. However, it is not clear whether DRG resums logarithmic corrections at all orders, as regular RG does. The bulk of the difficulty in addressing this problem lies in the evaluation of higher loop integrals. However, if we can extract the general properties of such loops, without doing the calculations explicitly (see for e.g.\ \eqref{eq:DivergenceStructure}), perhaps we could improve the underpinnings of DRG itself. Alternatively, the fact that the SdSET correctly reproduces IR divergences of the full theory can be taken as evidence that the DRG works. From that perspective, investigating higher loops in dS could further strengthen the validity of SdSET.
  \item When Mellin representations are used to simplify Feynman integrals in flat space, the variables $\{s_\ell\}$ serve as little more than auxiliary parameters to be integrated over. But in dS, these variables carry physical meaning, as the eigenvalues of the dilatation generator. Furthermore, as we noted in \secref{sec:DivergenceStructure}, multidimensional MB integrals can be computed with the method of residues using a geometrical procedure on the hyperplanes $U_i({\bm s}) = -n$. We used these facts to identify that certain secular behaviors originate from subhorizon dynamics. We are left with the distinct impression that there may be deeper physical insights hidden away in the geometry of the aforementioned planes.
\end{enumerate}

Mellin space is fertile ground for the study of correlation function in dS, as many authors have noted before. It is our hope that the observations in this work reinforce that view, and encourage further investigation of these tools.

\paragraph{Acknowledgements:}

We are grateful to Daniel Green and Tim Cohen for providing valuable guidance and encouragement throughout this project. We also thank Charlotte Sleight and Chia-Hsien Shen for comments on the manuscript, and John McGreevy and Manuel Loparco for insightful discussions. A.P.\ was supported by the US~Department of Energy under grant no.~DE-SC0019035 and no.~DE-SC0009919.

\appendix
\appendixpage
\addappheadtotoc

\section{Anomalous dimensions: a geometric analogy}
\label{sec:AnomGeometric}

We use the terms `IR divergence' and `secular growth' in the context of loop integrals throughout the main text. We treat these artifacts with dynamical RG, whereas regular RG in flat space is prescribed for UV divergences. The relationship between these ideas can be understood intuitively by a simple analogy. Consider the dS metric
\begin{equation}
	\d s^2 = - \d t^2 + a(t)^2 \d \x^2, \qquad a(t) = e^{Ht} ,
\end{equation}
where $H$ is the Hubble parameter. This metric defines an expanding spacetime, which we may visualize as shown in \figref{fig:dS1}. Imagine now, some arbitrary shape in space, which also expands with the scale factor $a(t)$. We can define the dimension of this object by counting the minimum number of $H^{-3}$ sized balls required to cover it. We will assume that this number, $N$, is related to $H$ as
\begin{equation}
	N \propto H^\Delta \label{eq:DimensionDefn}
\end{equation}
where $\Delta$ is the \textit{dimension} of the shape. For instance, if the shape were a straight line of length $a(t)|\x|$, we could cover it with $N_{\rm line} = \left( \frac{a(t) |\x|}{H^{-1}} \right)^1$ balls. Therefore, the dimension of a straight line is $\Delta_{\rm line} = 1$, which makes sense. Let us take the shape in \figref{fig:dS1} to be a line, but we imbue it with the property that as it expands it reveals more structure. This is depicted in \figref{fig:ExpandingLine}. How does this behavior affect the dimension of the line?
%
\begin{figure}[ht]
  \centering
  \begin{tikzpicture}

    \def\basel{7}
    \def\height{5}
    \def\theta{30} 
    \def\arrowpos{0}


    \draw[thick]
      ({-\basel/2-(\height*tan(\theta)/2)},{-\height/2}) --
      ++({5*tan(\theta)},\height) coordinate[pos=\arrowpos] (L1) --
      ++(\basel,0) coordinate[pos=\arrowpos] (T1) --
      ++(-{5*tan(\theta)},-\height) coordinate[pos=\arrowpos] (R1)
      -- cycle coordinate[pos=\arrowpos] (B1);


    \begin{scope}[shift={(-2.5,-1)},scale=0.3,rotate=30]
      \draw[thick] (0,0)
        \foreach \i in {1,...,20} { -- (\i,{sin(360*\i/10)}) };

      \draw[thick,red] (0,0) node[below=0.3cm] {$H^{-3}$} ellipse (0.8 cm and 0.4 cm);

      \foreach \i in {1,...,20} {
        \draw[thick,red] (\i,{sin(360*\i/10)}) ellipse (0.8 cm and 0.4 cm);
      }
    \end{scope}

    \draw[thick,dashed,shift={(0.35,-3)},scale=0.6]
      ({-\basel/2-(\height*tan(\theta)/2)},{-\height/2}) --
      ++({5*tan(\theta)},\height) coordinate[pos=\arrowpos] (L2) --
      ++(\basel,0) coordinate[pos=\arrowpos] (T2) --
      ++(-{5*tan(\theta)},-\height) coordinate[pos=\arrowpos] (R2)
      -- ({-\basel/2-(\height*tan(\theta)/2)},{-\height/2}) coordinate[pos=\arrowpos] (B2);

      \draw[thick,dashed,shift={(-1,-3.5)},scale=0.25*0.6,rotate=30] (0,0)
        \foreach \i in {1,...,20} { -- (\i,{sin(360*\i/10)}) };

    \draw[thick,-{stealth'},blue] (L2) -- (L1);
    \draw[thick,-{stealth'},blue] (T2) -- (T1);
    \draw[thick,-{stealth'},blue] (R2) -- (R1);
    \draw[thick,-{stealth'},blue] (B2) -- (B1);

    \draw[very thick,-{stealth'}] (4.5,-3) -- ++(0,3) node[pos=0.5,right] {time};

  \end{tikzpicture}
	\caption{\label{fig:dS1} A cartoon of de Sitter spacetime. Three dimensional space is visualized as a plane that expands over time. Also shown is a path covered with balls of fixed size $H^{-3}$, the characteristic length scale of a fixed dS manifold.}
\end{figure}

\begin{figure}
  \centering
  \begin{tikzpicture}[scale=0.64]

      \draw[thick,-{stealth'}] (2,11.5) -- ++(11,0) node[pos=0.5,above] {time};

      \begin{scope}[shift={(0,10-4/sqrt(2))},rotate=45]
        \draw[thick] (0,0) -- (5,0);

        \draw[shift={(0.1,-1.25)},{stealth'}-{stealth'}] (0,0) -- (5,0) node[pos=0.5,circle,draw=none,fill=white,sloped] {$a(t) x$};

        \foreach \i in {0,...,4}
          \draw[red] ({\i+0.5},0) circle (0.5cm);

        \node at (2.25,1.5) {$\Delta_{\rm line} = 1$};
      \end{scope}


      \begin{scope}[shift={(1.5,3.25)},scale=0.1]
        \foreach \x in {0,...,59}
          \draw[thick] (\x,\x) -- ++(0,1) -- ++(1,0);
      \end{scope}
      \begin{scope}[shift={(1.5,3.25)},rotate=45]
      \foreach \i in {0,...,8}
        \draw[red] ({0.92*\i+0.5},0) circle (0.5cm);
      \end{scope}

      \begin{scope}[shift={(3,-2.5)}]
        \foreach \x in {0,...,9}{
          \draw[thick] (\x,\x) -- ++(0,1) coordinate[pos=0.5] (v) -- ++(1,0) coordinate[pos=0.5] (h);
          \draw[red] (v) circle (0.5cm);
          \draw[red] (h) circle (0.5cm);
        }
        \draw[thick,dashed] (-1,0) -- (0,0);
        \draw[thick,dashed] (10,10) -- (10,11);
        \node at (7.25,4.25) {$\Delta_{\rm line} = 1 + \gamma$};
      \end{scope}
  \end{tikzpicture}
	\caption{\label{fig:ExpandingLine} A line in dS that reveals more structure as it expands, a consequence of which is a change in its dimensions.}
\end{figure}

Once again we cover the line with balls of size $H^{-3}$ and add up their number. Due to its newly revealed structure, we will require more balls to cover the line than if it were a perfectly straight line. If we still assume that $N$ scales with $H$ as in \eqref{eq:DimensionDefn}, we can get the new number of balls required by modifying the value of $\Delta_{\rm line} \to 1 + \gamma$,
\begin{equation}
	N_{\rm line} = \left( \frac{a(t) |\x|}{H^{-1}} \right)^{1+\gamma} . \label{eq:AnomDimDefn}
\end{equation}
The number $\gamma$ is called the \textit{anomalous dimension} of the expanding line, and it is smaller than 1. This captures the intuition that the jagged line is not a one-dimensional object anymore, but it is not quite a two-dimensional object either. Similar considerations apply to other shapes as well.

We can use this picture to understand anomalous dimensions of quantum fields in a dS background. Microscopic quantum fluctuations, which get stretched out by dS expansion, continually source `structure' to the field operators. This is reflected in the anomalous dimension of the operator, like the ones we compute in this paper. Furthermore, we may use the fact that $\gamma < 1$ to expand \eqref{eq:AnomDimDefn} to
\begin{equation}
	N_{\rm line} = \left( \frac{a(t) |\x|}{H^{-1}} \right)^{1}
		\left( 1 + \gamma \log (aH |\x|) + \cdots \right) .
\end{equation}
That is, if we were doing perturbative calculations in some parameter of size $O(\gamma)$, the appearance of $\log(aH |\x|)$ indicates anomalous scaling.

The analogy helps us clarify some of the terminology in the paper. First, $\log(aH |\x|)$ (or $\log(|\k|/aH)$) blows up as $aH |\x| \to \infty$ (or $|\k|/aH \to 0$). This is the sense in which we refer to correlators containing these terms as IR divergent. Second, these terms arise from the time evolution of subhorizon modes to superhorizon scales, which is why we also call them secular growth terms\footnote{This is not the only source of secular growth -- it can also arise from the classical evolution of the superhorizon modes. See, for example, (3.9) from \cite{Cohen:2021fzf}.}. Third, it is really the UV details that contribute to the anomalous scaling at late time. In a flat space loop calculation the appearance of a $\log(\mu |\x|)$, where $\mu$ is some renormalization scale, signals exactly the same thing: the UV structure of the theory affects the scaling of the operators at long wavelength. The parallel between dynamical RG and regular RG should now be apparent.



\phantomsection
\addcontentsline{toc}{section}{References}
\small
\bibliographystyle{utphys}
\input{LoopsIndS.bbl}

\end{document}

%% file: LoopsIndS.bbl
\providecommand{\href}[2]{#2}\begingroup\raggedright\endgroup